\documentclass[amsmath,amssymb,onecolumn,superscriptaddress]{revtex4}

\usepackage{graphicx}% 
\usepackage{dcolumn}% 
\def\be{\begin{equation}}
\def\ee{\end{equation}}
\def\beq{\begin{eqnarray}}
\def\eeq{\end{eqnarray}}

\usepackage{bm}
\usepackage{graphicx}
\usepackage{dcolumn}
\usepackage{amsmath}
\usepackage[latin1]{inputenc}
\usepackage{graphicx, psfrag}
\usepackage{amssymb}
\usepackage[colorlinks=true, citecolor=blue, urlcolor = blue, linkcolor= red, bookmarks=true]{hyperref}
\usepackage{float}
\usepackage{mathtools}
\usepackage{amsmath}
\usepackage{amsfonts}
\usepackage{dcolumn}
\usepackage{hyperref}
\usepackage{subfigure}
\usepackage{pgfplots}
\usepackage{epstopdf}
\usepackage{booktabs}
\usepackage{orcidlink}
\usepackage{xcolor}

\begin{document}
\title{Charged strange star coupled to anisotropic dark energy in Tolman-Kuchowicz spacetime}

\author{Pramit Rej \orcidlink{0000-0001-5359-0655} \footnote{Corresponding author}}
\email[Email:]{pramitrej@gmail.com, pramitr@sccollegednk.ac.in }
 \affiliation{Department of Mathematics, Sarat Centenary College, Dhaniakhali, Hooghly, West Bengal 712 302, India}

\author{Akashdip Karmakar \orcidlink{0009-0007-3848-1443}}
\email[Email:]{akashdip999@gmail.com}
 \affiliation{Department of Mathematics,  Indian Institute of Engineering Science and Technology, Shibpur, Howrah, West Bengal 711 103, India}

\begin{abstract}\noindent
The concept of dark energy can be used as a possible option to prevent the gravitational collapse of compact objects into singularities. It affects the universe on the largest scale, as it is responsible for our universe's accelerated expansion. As a consequence, it seems possible that dark energy will interact with any compact astrophysical stellar object [Phys. Rev. D 103, 084042 (2021)]. In this work, our prime focus is to develop a simplified model of a charged strange star coupled to anisotropic dark energy in Tolman-Kuchowicz spacetime (Tolman, Phys Rev 55:364, 1939; Kuchowicz, Acta Phys Pol 33:541, 1968) within the context of general relativity. To develop our model, here we consider a particular strange star object, Her X-1 with observed values of mass $=(0.85 \pm 0.15)M_{\odot}$ and radius $= 8.1_{-0.41}^{+0.41}$ km. respectively. In this context, we initially started with the equation of state (EoS) to model the dark energy, in which the dark energy density is proportional to the isotropic perfect fluid matter-energy density. The unknown constants present in the metric have been calculated by using the Darmois-Israel condition. We perform an in-depth analysis of the stability and force equilibrium of our proposed stellar configuration as well as multiple physical attributes of the model such as metric function, pressure, density, mass-radius relation, and dark energy parameters by varying dark energy coupling parameter $\alpha$. Thus after a thorough theoretical analysis, we found that our proposed model is free from any singularity and also satisfies all stability criteria to be a stable and physically realistic stellar model.

\end{abstract}

\maketitle

\section{Introduction}
The general theory of relativity (GTR), developed by Albert Einstein, is a significant gravitational instrument for understanding the fabric of space-time as well as the dynamics of cosmic bodies and other phenomena.
Due to the challenges for obtaining the exact analytic solutions of Einstein field equations describing compact configurations such as strange stars, the theoretical studies modeling have significantly improved, in particular designing fluid sphere models with anisotropic matter distributions  
i.e. unequal radial and tangential pressure: $p_r \neq p_t.$ Anisotropies play a significant role in the stability and equilibrium of the stellar structure
 \cite{Maurya:2019zyc}.
 There is a wealth of research on the impact of local anisotropy on the global characteristics of relativistic compact objects \cite{tamta2017new,bondi1992anisotropic, DEB2017239, dev2002anisotropic,di1997cracking, krori1984some,mak2003anisotropic, maurya2016new}.\\
Now it is quite an interesting fact that according to cosmology, the universe's apparent matter accounts for only $5\%$ of gravity, with the rest $26\%$ and $69\%$ being explained by dark matter and dark energy, respectively.  
Dark energy is considered a spatially homogeneous cosmic fluid, although it can be developed to non-homogeneous spherically symmetric spacetimes by assuming that the pressure in the dark energy equation of state is negative radial pressure and that the transverse pressure can be obtained using the field equations \cite{Lobo:2005uf}. The significance of dark energy is not diminished as the universe expands because it is evenly distributed throughout the universe, both in space and in time. The Supernova Cosmology Project at the Lawrence Berkeley National Laboratory and the High-z Supernova Search Team made observations of type-Ia supernovae ("one-A") in 1998 \cite{SupernovaCosmologyProject:1998vns}, which led them to hypothesize that the universe is expanding in accelerating manner. These type-Ia supernovae provide the most concrete evidence for dark energy. 
Numerous studies have been conducted on compact astrophysical objects whose interior pressure $p$ and energy density $\rho$ obey an equation of state typical of dark energy, such as $p = -\rho$. In the literature, these objects have been given a variety of names. For ease of use, we call them "dark energy stars"\cite{beltracchi2019formation, Chapline:2004jfp}. Perhaps dark energy's origins remain a mystery but by observing how quickly the universe is expanding and how quickly large-scale structures like galaxies and clusters of galaxies develop as a result of gravitational instabilities, one can determine the presence of dark energy.
Even though dark energy stars have a chance to exhibit spacetime singularities, non-singular dark energy stars are more commonly of interest. Astrophysical measurements made recently have shown that the universe is expanding faster than before. Observationally by using the Hubble Space Telescope, Riess and co-researchers discovered that the universe is expanding $9 \%$ faster than predicted \cite{Riess:2019cxk}.
Although it works against gravity, dark energy speeds up the universe's expansion and slows down the development of large-scale structures which leads to the violation of the strong energy condition. We naturally seek local astrophysical appearances of dark energy because of the fundamental significance of the phenomenon in cosmology. The equation of state of dark energy may be explained by $p = \omega\rho$, with $\omega < -\frac{1}{3}$ \cite{sushkov2005wormholes,bibi2016solution}. In this connection, Feng et al. examined different phenomenological interaction models for dark energy and dark matter by performing statistical joint analysis with three different observational data \cite{feng2008observational}\\
Recently, novel findings of neutron stars and strange stars that meet the exact solutions of the 4-D Einstein field equations have been made possible by astronomical studies of compact objects. Many researchers have created a large number of precise models using the Einstein-Maxwell field equations \cite{gupta2011class,kiess2012exact,takisa2013some,malaver2017new,malaver2018generalized,sunzu2014quark}. 
The most significant and logical explanations for the occurrence of tangential pressures inside a star are the existence of a solid core, the presence of type 3A superfluid, magnetic field, phase transitions, a pion condensation, and electric field \cite{usov2004electric}.
Anisotropic pressures are present in several astronomical objects, including the X-ray pulsar, Her X-1, 4U 1820-30, and SAX J1804.4-3658. In order to have a minimal impact on the structure of the star, it is true that all macroscopic entities are charged or that they may have a minor amount of charge, as proposed by Glendening \cite{glendenning2012compact}. So, as it is possible to claim that stable charged astrophysical compact objects exist in nature, we have chosen to study the charged case. The electrostatic repulsion of the Coulomb force and the outward pressure gradient will balance the gravitational contraction in the presence of charge. In this scenario, a point singularity caused by the gravitational collapse of a spherically symmetric matter distribution with charge may be removed \cite{das2015anisotropic}. Ray et al. studied the effect of electric charge in compact stars assuming that the charge distribution is proportional to the mass density and the relativistic hydrostatic equilibrium equation, i.e., the Tolman-Oppenheimer-Volkoff equation, gets modified due to the inclusion of electric charge \cite{ray2003electrically}. The dynamics of cold star models, compact stars, strange stars, and other stellar configurations with the hybrid coupling of strange matter have been the subject of continuous research during the last decade \cite{pant2011some, pant2011well, pant2019core, pant2020three, singh2017physical, gedela2019relativistic, mafa2013compact, ngubelanga2015ray, matondo2018relativistic, singh2020compact, rahaman2012singularity, maharaj2014some, estevez2020quintessence, estevez2020tolman}.
\\
$~~~~~$ On the other hand recently many models of particular dark energy stars have been put out in literature \cite{chan2009star, chan2009anisotropic, lobo2007gravastars, chan2011gravastars, bertolami2005chaplygin, cattoen2005visser}.
Bhar and her co-workers have already suggested a model that might be helpful in examining the possible clustering of dark energy and also very recently proposed a model for a dark energy star made up of dark and ordinary matter in which the density of dark energy is proportional to the density of isotropic perfect fluid matter. In-depth research was done on the stability of stellar configuration as well as the model's physical characteristics such as pressure, density, mass function, and surface redshift \cite{bhar2018dark, bhar2021dark}.
\\
In this paper, we provide a model for a charged strange star coupled to inhomogeneous anisotropic dark energy in Tolman-Kuchowicz 
spacetime. Here, we made the assumption that the isotropic perfect fluid matter density is proportional to the radial pressure that dark energy exerts on the system. Here, in this case of Einstein-Maxwell field equation one has eight unknown functions i.e.  
\{$\rho$, $\rho^{de}$, $p$, $p^{de}_r$, $p^{de}_t$, $\nu$, $\lambda$, $E$\} and four equations. Therefore, in order to solve this system it is necessary to give additional information or conditions that have been discussed in our work. We have described the fundamental field equations for a charged strange star model when inhomogeneous anisotropic dark energy is present in the system. There is a detailed picture of our used metric coefficients. The solutions to the field equations and a smooth correspondence between internal and external spacetime are discussed. A rigorous discussion has been made regarding the physical study of our current model. Also, we summarise some key findings on the formation of a dense, stable stellar structure by the dark energy's repulsive scalar field.\par

Our present article is designed as follows. It has five core sections.  We have described the technique to solve Einstein-Maxwell field equations for charged static and spherically symmetric matter distribution coupled with dark energy by employing the well-known Tolman-Kuchowicz Metric in section \ref{interior} and we fix the constants of our model using junction conditions for a particular compact star candidate Her X - 1. We have analyzed the physical attributes of our model in the next section \ref{phy}. We perform the stability analysis through different aspects and check the equilibrium of forces for our present model in Section \ref{stable}. In the final section \ref{con}, the key findings of our current investigation have been discussed with some concluding remarks.

\section{Solutions of Einstein-Maxwell Field Equations}\label{interior}
In order to describe the interior space-time of a static spherically symmetric compact object, here we consider the interior line element in a static spherically symmetric $4$D space-time in the Schwarzchild coordinate system $(x^i=t,\,r,\,\theta,\,\phi)$ as follows,
\begin{equation} \label{line1}
ds^{2}=e^{\nu (r)}dt^{2}-e^{\lambda (r)}dr^{2}-r^{2}(d\theta^{2}+\sin^{2}\theta d\phi^{2}),
\end{equation}
Let us assume that the energy-momentum tensor for an anisotropic charged with two fluids is composed of matter with matter-energy density $\rho$, pressure $p$ of the corresponding baryonic matter, electric field intensity $E$, and 'dark' energy density $\rho^{de}$, corresponding radial pressure $p_r^{de}$, and tangential pressure $p_t^{de}$ respectively. The sub-index
"de" stands for dark energy throughout the paper. This dark energy density can be expressed in terms of the (variable) cosmological constant ($\Lambda$) as $\rho^{\text{de}}=\frac{\Lambda}{8\pi}$ \cite{Ghezzi:2009ct}.\\
We can write the corresponding energy-momentum tensor for an anisotropic charged with two fluids \cite{Ghezzi:2009ct} as,
 \begin{equation} \label{t1}
  \begin{rcases}
    \begin{aligned}
      T_0^0 &=\rho^{\text{eff}}+E^2= (\rho+\rho^{de}) +E^2, \\
      T_1^1 &=-p_r^{\text{eff}}+E^2=-(p+p_r^{de}) +E^2, \\
      T_2^2 &=T_3^3=-p_t^{\text{eff}}-E^2=-(p+p_t^{de}) -E^2 ,\\
      T_0^1 &=T_1^0=0.
    \end{aligned}
  \end{rcases} \text{Energy-Momentum Tensor}
\end{equation}
where, $\rho^{\text{eff}}=(\rho+\rho^{de}$), $p_r^{\text{eff}}=(p+p_r^{de}$), and $p_t^{\text{eff}}=(p+p_t^{de}$) are the effective energy density and pressure components respectively.\par

Now assuming geometricized unit system ($G = c = 1$), the Einstein-Maxwell field equations are given as follows,
\begin{eqnarray}
T_0^0:~\kappa(\rho+\rho^{de}) +E^2 &=&e^{-\lambda}\left(\frac{\lambda'}{r}-\frac{1}{r^{2}} \right)+\frac{1}{r^{2}}, \label{fe1}\\
T_1^1:~\kappa (p+p_r^{de}) -E^2  &=&e^{-\lambda}\left(\frac{1}{r^{2}}+\frac{\nu'}{r} \right)-\frac{1}{r^{2}}, \label{fe2}\\
T_2^2=T_3^3:~\kappa (p+p_t^{de}) +E^2 &=& \frac{1}{2}e^{-\lambda}\left( \frac{\nu'^{2}}{2}+\nu''-\frac{\lambda'\nu'}{2}+\frac{\nu'-\lambda'}{r}\right)\label{fe3}
\end{eqnarray}
where primes $(^{\prime})$ denotes the derivative with respect to the radial coordinate `r' and $\kappa=8\pi$.\par
Here $E$ is the electric field intensity, $\sigma=\sigma(r)$ is the proper charge density connected by the relation,
\begin{eqnarray}
\kappa \sigma &=& \frac{2}{r^2} e^{-\lambda /2} (r^2 E)',\label{fe4}
\end{eqnarray}
Equation (\ref{fe4}) can also be equivalently expressed as,
\begin{eqnarray}
E(r) &=& \frac{\kappa}{2r^2} \int_0^r \sigma r^2 e^{\lambda /2}  dr = \frac{q(r)}{r^2},\label{fe5}
\end{eqnarray}
where $q(r)$ represents the total charge contained within the sphere of radius $r$ under consideration.

For our present paper we have taken the well-known metric potentials proposed by Tolman-Kuchowicz \cite{Tolman:1939jz,osti_4507306} (henceforth TK {\em ansatz}) as,
\begin{equation}\label{tk}
  \begin{rcases}
    \begin{aligned}
\nu(r)= Br^2+2\ln D,\\
\lambda(r)= \ln(1 + ar^2 + br^4)
\end{aligned}
  \end{rcases} \text{Tolman-Kuchowicz Metric}
\end{equation}
where $a$, $B$ and $b$ are constants with units km$^{-2}$, km$^{-2}$ and km$^{-4}$, respectively. $D$ is a dimensionless constant. We shall calculate the numerical values of these constants from a smooth matching of interior and exterior spacetimes. The metric potentials provide a non-singular stellar model which will be described in the coming sections.\par

With the help of the expressions given in (\ref{tk}), the field equations (\ref{fe1})-(\ref{fe3}) take the following form :
\begin{eqnarray}
\kappa(\rho+\rho^{de}) +E^2 &=&\frac{3 a + (a^2 + 5 b) r^2 + 2 a b r^4 +
 b^2 r^6}{(1 + a r^2 + b r^4)^2},   \label{f1}\\
\kappa (p+p_r^{de}) - E^2 &=& \frac{2B-a -b r^2}{1 + a r^2 + b r^4}, \label{f2}\\
\kappa (p+p_t^{de}) +E^2 &=&\frac{2B - a + \big( B (a + B)-2 b\big) r^2 + a B^2 r^4 +
 b B^2 r^6}{(1 + a r^2 + b r^4)^2} .\label{f3}
\end{eqnarray}

\subsection{Assumption of Equation of State due to the dark energy}\label{eosde}

To obtain the explicit solutions of the above field equations (\ref{f1})-(\ref{f3}) is too much complicated due to its non-linearity nature, although the system of equations is mathematically well defined. So to remove this complicacy, we have adopted three assumptions as proposed earlier by the authors mentioned in the Refs. \cite{Ghezzi:2005iy,Ghezzi:2009ct,Barreto:2006cr} given as follows:\\

(i) The radial dark pressure ($p_r^{de}$) is proportional to the density corresponding to dark energy, i.e.,
\begin{eqnarray}\label{as1}
p_r^{de}=-\rho^{de},
\end{eqnarray}

(ii) The density corresponding to dark energy ($\rho^{de}$) is proportional to the normal baryonic matter density, i.e.,
\begin{eqnarray}\label{as2}
\rho^{de}=\alpha \rho,
\end{eqnarray}
where $\alpha(>0)$ is a proportionality constant. Here
we called it "coupling parameter". \\

Clearly, we see that this type of equation of state corresponds to dark energy and is similar to the MIT Bag model for hadrons \cite{glendenning2012compact}. The above dark energy EoS relates to the MIT Bag model's Bag constant contribution. {\bf Here we assume the Bag constant to be density-dependent. In astrophysical models of hybrid stars with a deconfined phase of quarks at the core, the MIT Bag model is employed.}\\

(iii) lastly, upon following these earlier works in Refs. \cite{Barreto:2006cr, das2015anisotropic} we preferably assume the difference between radial and tangential pressure corresponding to dark energy is proportional to the square of the electric field intensity $E$, i.e.
\begin{eqnarray}\label{as3}
 p_t^{de} - p_r^{de} = \frac{E^2}{4\pi}  
\end{eqnarray}
\subsection{Proposed model of dark energy star}
Now solving the equations (\ref{f1})-(\ref{f3}) by the help of these above assumptions (\ref{as1})-(\ref{as3}), we obtain the explicit expressions for $\rho$, $p$ and $E^2$ as,
\begin{eqnarray}
 \rho &=& \frac{12a + (3 a^2 + 21 b + a B - B^2)r^2 + (6 a b + 2 b B - a B^2)r^4 + b (3 b - B^2)r^6}{4\kappa (1 + \alpha){(1 + ar^2 + br^4)}^2} ,   \label{f11}\\
 p &=&  \frac{1}{4\kappa(1 + \alpha){(1 + ar^2 + br^4)}^2}\Bigg[a (-4 + 8 \alpha) + 
 8 (1 + \alpha) B + \big(-3 a^2 + (-5 + 16 \alpha) b + a (7 + 8 \alpha) B + B^2 \big) r^2 \nonumber\\ && + \big(2 (3 + 4 \alpha) b B + a (-6 b + B^2)\big) r^4 +  b (-3 b + B^2) r^6 \Bigg]   ,\label{f12}\\
E^2 &=&  \frac{r^2 \Big[a^2 - a B + B^2 + a (2 b + B^2) r^2 + 
   b \Big(-1 - 2 B r^2 + (b + B^2) r^4\Big)\Big]}{4 (1 + a r^2 + b r^4)^2} .\label{f13}
\end{eqnarray}
The charge density ($\sigma$) is obtained from Eqn. (\ref{fe4})as,
\begin{eqnarray} \label{sig}
\sigma &=& \frac{(r^2 E)^{'}}{4\pi r^2\sqrt{e^\lambda}}
\nonumber\\ 
&=& \frac{1}{\kappa(1 + ar^2 + br^4)^{\frac{7}{2}} \sqrt{\frac{r^2[a^2 - aB + B^2 + a(2b + B^2)r^2 + b\{-1 - 2Br^2 + (b + B^2)r^4\}]}{(1 + ar^2 + br^4)^2}}}\Bigg[3(a^2 - b - aB + B^2)r \nonumber\\ && \
+ (a^3 + 7ab - {a^2}B - 8bB + 5aB^2)r^3 + \{3b(a^2 + 2b) - 3abB + 2(a^2 + 2b)B^2\}r^5 \nonumber\\ && \ + 3ab(b + B^2)r^7 + b^2(b + B^2)r^9\Bigg]
\end{eqnarray}
The expressions for matter-energy density, radial and transverse pressure due to the dark energy are obtained as,
\begin{eqnarray} 
\rho^{de} &=&  \frac{\alpha [12a + (3 a^2 + 21 b + a B - B^2)r^2 + (6 a b + 2 b B - a B^2)r^4 + b (3 b - B^2)r^6]}{4\kappa (1 + \alpha){(1 + ar^2 + br^4)}^2},    \label{de1}\\
p_r^{de} &=& - \frac{\alpha [12a + (3 a^2 + 21 b + a B - B^2)r^2 + (6 a b + 2 b B - a B^2)r^4 + b (3 b - B^2)r^6]}{4\kappa (1 + \alpha){(1 + ar^2 + br^4)}^2},   \label{de2}\\
p_t^{de}  &=&  \frac{1}{4\kappa (1 + \alpha){(1 + ar^2 + br^4)}^2}\Bigg[-12a\alpha - \{a^2(-2 + \alpha) + (2 + 23\alpha)b + a(2 + 3\alpha)B - (2 + 3\alpha)B^2\}r^2 + \nonumber\\ && \{-2a(-2 + \alpha)b - 2(2 + 3\alpha)bB + a(2 + 3\alpha)B^2\}r^4 + b\{-(-2 + \alpha)b + (2 + 3\alpha)B^2\}r^6\Bigg].  \label{de3}
\end{eqnarray}

The expressions for effective matter-energy density, effective radial and transverse pressure for our present model are obtained as,

\begin{eqnarray} 
\rho^{\text{eff}} &=& \rho + \rho^{de}   \nonumber\\ 
&=& \frac{12a + (3a^2 + 21b + aB - B^2)r^2 + (6ab + 2bB - aB^2)r^4 + b(3b - B^2)r^6}{4\kappa{(1 + ar^2 + br^4)}^2}, \label{ef1}\\ \nonumber\\
p_r^{\text{eff}} &=& p + p_r^{de}  \nonumber\\ 
&=& \frac{-4a + 8B + (-3a^2 -5b + 7aB + B^2)r^2 + \{6bB + a(-6b +B^2)\}r^4  + b(-3b + B^2)r^6}{4\kappa {(1 + ar^2 + br^4)}^2},\label{ef2}\\ \nonumber\\
p_t^{\text{eff}} &=& p + p_t^{de}  \nonumber\\ 
&=& \frac{-4 a + 8 B - (a^2 + 7 b - 5 a B - 3 B^2) r^2 - (2 a b - 2 b B - 
    3 a B^2) r^4 - b (b - 3 B^2) r^6}{4\kappa {(1 + ar^2 + br^4)}^2}.  \label{ef3}
\end{eqnarray}

\subsection{Determination of constants using Junction or Matching Conditions}\label{match}

In order to develop the profiles of the model parameters, the values of $a$, $b$, $B$, and $D$ must be fixed. So as to investigate the significant values of unknown constants, we match our interior space-time smoothly to the exterior space-time described by the Reissner-Nordstr\"om metric \cite{reissner1916eigengravitation,nordstrom1918energy} given by
\begin{eqnarray}
ds_{\Sigma}^{2} &=& -\left(1 - \frac{2M}{r} + \frac {Q^2}{r^2}\right)dt^2 + \left(1 - \frac{2M}{r} + \frac {Q^2}{r^2}\right)^{-1}dr^2
 + r^2(d\theta^2+\sin^2\theta d\phi^2), \label{eq22}
\end{eqnarray}
where $Q$ is the total charge enclosed within the surface $r = R$. 
Now the continuity of the metric coefficients $g_{tt}$, $g_{rr}$ and $\frac{\partial g_{tt}}{\partial r}$ across the boundary surface $r= R$ between the interior and the exterior regions give the following set of relations:
\begin{eqnarray}
1 - 2\tilde{\mathcal{X}}+ \tilde{\mathcal{Y}} = e^{\nu (R)} &=& D^2 e^{BR^2},\label{eq23}\\
1 - 2\tilde{\mathcal{X}} + \tilde{\mathcal{Y}} = e^{-\lambda (R)} &=& (1 + a R^2 + b R^4)^{-1},\label{eq24}\\
\tilde{\mathcal{X}} - \tilde{\mathcal{Y}} = R\Big[ \frac{\partial}{\partial r} e^{\nu (r)} \Big]_{r=R} &=& B R^2 D^2 e^{BR^2}.\label{eq25}
\end{eqnarray}
where, $\tilde{\mathcal{X}}=\frac{M}{R}$ and $\tilde{\mathcal{Y}}=\frac {Q^2}{R^2} $ and both $\tilde{\mathcal{X}},\,\tilde{\mathcal{Y}}$ are dimensionless quantity.\\
Solving the Eqs.~(\ref{eq23})-(\ref{eq25}) along with the  condition $p(r=R)=0$ , we determine the values of the constants $B$, $D$, $a$ and $b$ as,
\begin{eqnarray}
B &=& \frac{M R -Q^2  }{R^2 \Big[Q^2 + R (R-2 M )\Big]},\label{eq26}\\ 
D &=&  e^{-BR^2 /2}\sqrt{1 - 2\tilde{\mathcal{X}}+ \tilde{\mathcal{Y}}}, \label{eq27}\\ 
a &=& -\frac{1}{R^2} - b R^2 + \frac{1}{Q^2 - 2 M R + R^2}, \label{eq28}\\ 
b &=& \frac{1}{R^4 (1 - 8 \alpha + B R^2) \Big[Q^2 + R (R-2 M )\Big]^2}\Bigg[Q^4 (1 - 8 \alpha + B R^2) + 
 Q^2 R \Big\{4 R - 8 \alpha R + (9 + 8 \alpha) B R^3 + B^2 R^5 \nonumber\\&& + 4 M (-1 + 8 \alpha - B R^2)\Big\} + 
 R^2 \Big\{4 M^2 (1 - 8 \alpha + B R^2) + B R^4 (8 + 8 \alpha + B R^2) - 
    2 M R \Big(4 + 8 \alpha (-1 + B R^2)\nonumber\\&&  + B R^2 (9 + B R^2)\Big)\Big\}\Bigg]\label{eq29}
\end{eqnarray} 
Thus, we have successfully obtained the values of all constants present in our model in terms of $M, R$, and $Q$. From (\ref{eq28}) and (\ref{eq29}) we see that $a$ and $b$ are functions of dark energy coupling parameter $\alpha$.
Now to analyze the physical attributes of our present model we have considered here particularly the compact star Her X - 1 with observed mass and radius $M = 0.85 \pm 0.15~M_{\odot},\, R = 8.1_{-0.41}^{+0.41}$ km \cite{Abubekerov:2008inw}. Along with this, we have also assumed $Q = 1.31$. The reason behind this particular choice is that according to earlier work by Varela et al. \cite{Varela:2010mf}, the square of the charge-radius ratio $\big(\frac{Q^2}{R^2}\big)$ for a charged anisotropic fluid sphere should lie within the interval $[0, 0.543)$. For our chosen compact object HER X-1, the value is $\frac{Q^2}{R^2} = 0.0237522$ (Approx.), i.e. within the proposed range, and thus we strengthen our numerical calculations as well as graphical analysis.

\section{Analysis of stellar physical attributes}\label{phy}

\subsection{Regularity of metric potentials}\label{mp}
\begin{figure}[H]
    \centering
        \includegraphics[scale=.55]{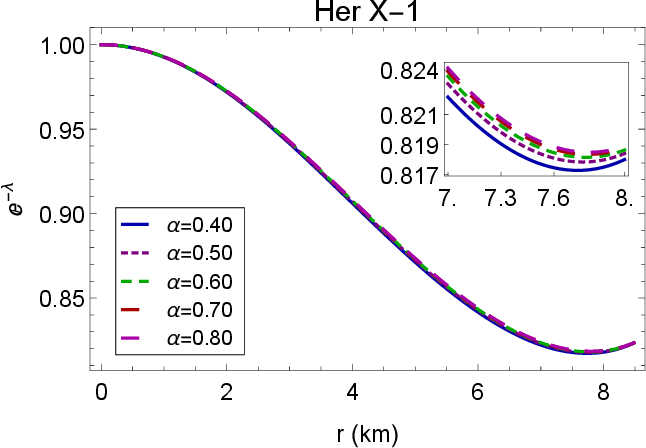}
         \includegraphics[scale=.55]{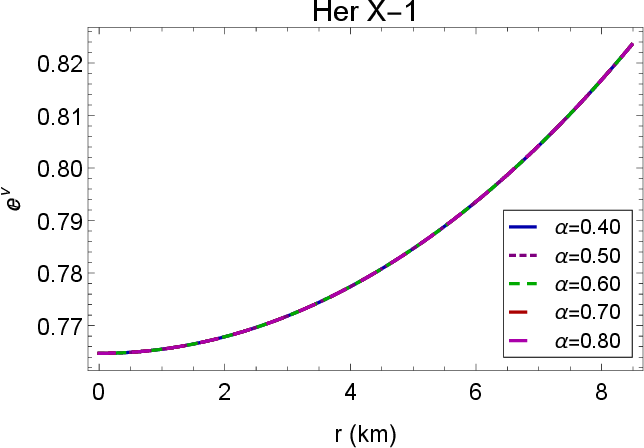}
        \caption{Variation of metric functions $e^{-\lambda(r)}$ and $e^{\nu(r)}$ with respect to `r' with required magnified inset.}\label{metric}
\end{figure}

Here we have analyzed the behavior of the metric potential temporal components $e^{\nu(r)}$ and spatial components $e^{\lambda(r)}$.
We can easily check that, $[{e^{\nu(r)}}]_{r = 0}= D^2$, a non-zero constant, and $[{e^{-\lambda(r)}}]_{r=0} = 1$ and which lead to the fact that both metric potential components are finite at the centre and having regularity at all points, $r < R$ \cite{Delgaty:1998uy, Pant:2010iub}. Moreover, $\Big[\frac{d(e^{\nu(r)})}{dr}\Big]_{r=0} = (2B{D^2}r{e^{Br^2}})\Big\rvert_{r=0} =0$ \text{and}\\ $\Big[\frac{d(e^{\lambda(r)})}{dr}\Big]_{r=0} = (2ar + 4br^3)\Big\rvert_{r=0} =0$,\\
i.e. the derivatives of the metric potential components vanish at the center of the star. They are also positive and consistent within the interior of the star which can be easily verified from the radial profiles of the metric coefficient components shown in Fig.~\ref{metric}. Here we have shown the regularity of the metric potentials with varying coupling parameters $\alpha$. The internal space-time smoothly matches with the asymptotically flat exterior space-time at the surface ($r = R$), fulfilling the Darmois-Israel condition \cite{Chu:2021uec, darmois1927equations, Israel:1966rt}. By matching these, we obtain the values of the constant parameters that characterize our model. Thus we verify that the metric potential components are well-behaved within the range $(0, R)$.

\subsection{Stellar profiles of baryonic matter-energy density, pressure, and electric field}

\begin{figure}[H]
    \centering
        \includegraphics[scale=.47]{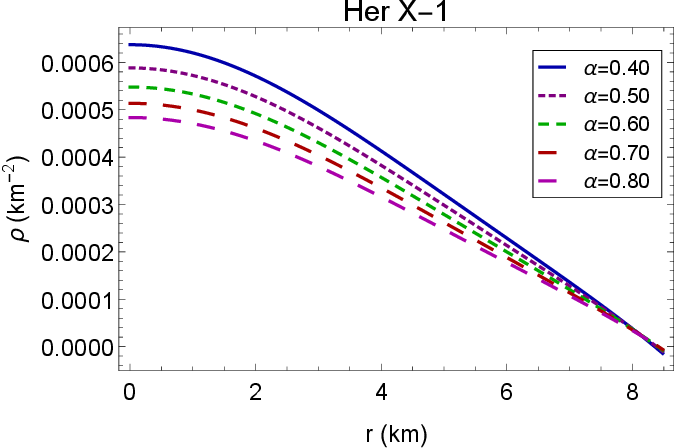}
         \includegraphics[scale=.47]{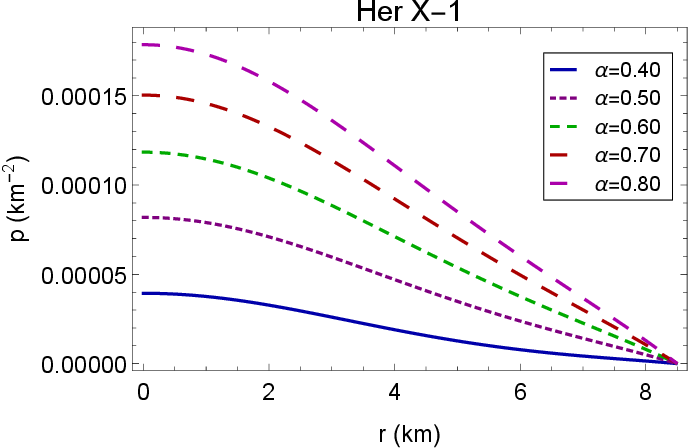}
         \includegraphics[scale=.47]{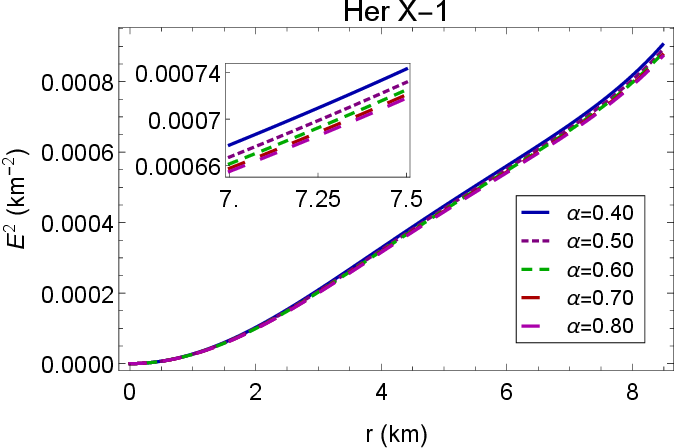}
        \caption{Profiles of baryonic matter-energy density, pressure, and electric field intensity ($E^2$) with respect to `r' with required magnified inset.}\label{rho}
\end{figure}

The density of the confined matter plays a key role in determining how stable a star structure is against gravitational collapse and similarly pressure plays a significant role in defining the stellar boundary and overall stability \cite{chandrasekhar1984stars}. From equations (\ref{f11}) and (\ref{f12}), we get the central values of $\rho~ \text{and}~ p$ as,
$\rho_c= [\rho]_{r=0} = \frac{3a}{\kappa (1 + \alpha)} > 0$,
 $p_c= [p]_{r=0} =  \frac{a(-1 + 2 \alpha) + 2 (1 + \alpha) B }{\kappa(1 + \alpha)} >0$. 
  Fig.~\ref{rho} indicates that matter-energy density and pressure are monotonic decreasing functions of radius $r$, having the maximum value at the center of the star. $[E^2]_{r=0} = 0.$ i.e. electric field intensity vanishes at the center and gradually increases when $r$ increases as shown in Fig.~\ref{rho}. Finally, it is also observed that $\rho$, $p$, and $E^2$ inside the star are non-negative. We have represented the numerical values of central density ($\rho_c$), surface density ($\rho_s$), central pressure ($p_c$) and the central pressure-density ratio $p_c/\rho_c$ in Table~\ref{tab1} for different values of the dark energy coupling parameter $\alpha$.

\begin{figure}[htbp]
    \centering
        \includegraphics[scale=.55]{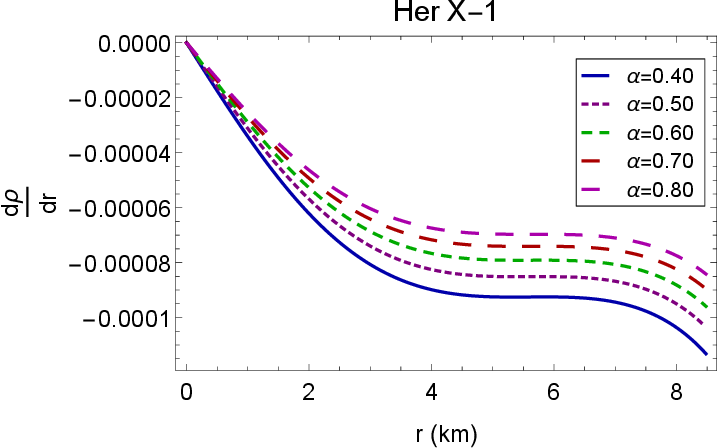}
         \includegraphics[scale=.55]{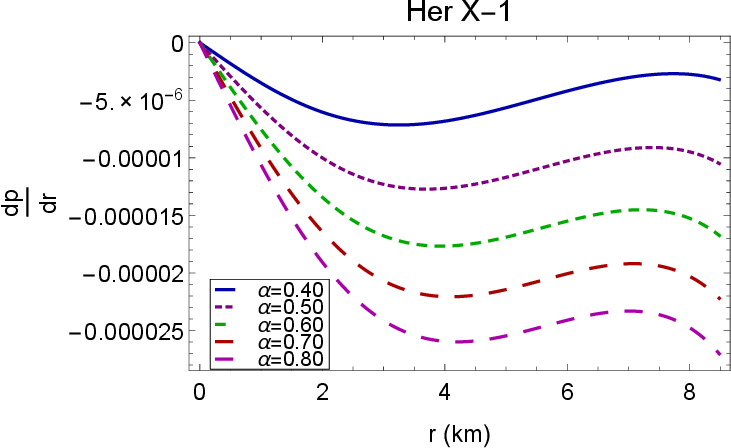}
        \caption{Gradients of matter density and pressure are plotted `r'.}\label{grad}
\end{figure}

Furthermore, the density and pressure gradients of the normal baryonic
matter for our present model are obtained as,

\begin{eqnarray}
  \frac{d\rho}{dr} &=& \frac{r}{2\kappa(1 + \alpha)(1 + ar^2 + br^4)^3}\big[-21a^2 + 21b + aB - B^2 - \{3a^3 + a^2B -4bB + a(57b + B^2)\}r^2 \nonumber\\ && - 3b\{18b + a(3a + B)\}r^4 + b\{-4bB + a(-9b + B^2)\}r^6 + b^2(-3b + B^2)r^8\big], 
\end{eqnarray}

\begin{eqnarray}
 \frac{dp}{dr} &=& \frac{r}{2\kappa(1 + \alpha)(1 + ar^2 + br^4)^3}\big[B^2 + 3a^3 r^2 + a^2\{5 - 16\alpha -(7 + 8\alpha)Br^2 + 9br^4\} \nonumber\\ && +a\{-(9 + 8\alpha)B +((9 - 48\alpha)b + B^2))r^2 - 3(7 + 8\alpha)bBr^4 + b(9b - B^2)r^6\} + b\{-5 + 16\alpha - 4(5 + 4\alpha)Br^2 \nonumber\\ && + 6(1- 8\alpha)br^4 - 4(3 + 4\alpha)bBr^6 + b(3b - B^2)r^8\} \big]  
\end{eqnarray}
From Fig.~\ref{grad}, we see that the density and pressure gradients stay negative throughout the fluid sphere and vanish at the core.
Also, taking second derivative of $\rho, p$ we get,
$$\Big[\frac{d^2\rho}{dr^2}\Big]_{r=0} = -\frac{21a^2 -21b -aB + B^2}{2\kappa(1 + \alpha)},$$\\
$$\Big[\frac{d^2 p}{dr^2}\Big]_{r=0} = -\frac{a^2(\alpha - 5) + (16 \alpha -5)b + a(9 + 8\alpha)B - B^2}{2\kappa(1 + \alpha)}$$
Clearly, the second derivatives take negative values at the center for the dark energy coupling parameter range, $\alpha \in [0.40, 0.80]$ that we have considered in this investigation and hence we can conclude that both the density and pressure take the maximum value at the core of the stellar configuration.

\subsection{Dark energy density and dark pressure}

In physical cosmology, dark energy is a fictitious type of energy that fills all of space and has the tendency to accelerate the universe's expansion \cite{Peebles:2002gy}. It is termed as "dark" because it does not have an electric charge and does not interact with electromagnetic radiation such as light. Dark energy has the reverse effect of positive energy in that it accelerates the expansion of the universe. Dark energy currently occupies over three-quarters of the total mass energy of the universe, according to the conventional cosmological model. Dark energy has recently been employed as an important factor in an attempt \cite{Baum:2006ee} to develop a cyclic model of the universe. Unusually, dark energy causes the universe's expansion because it has very strong negative pressure. This negative pressure induces gravitational repulsion.\par
\begin{figure}[H]
    \centering
        \includegraphics[scale=.46]{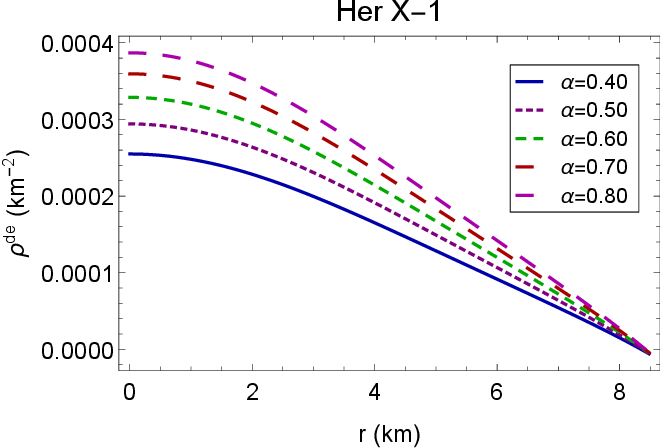}
        \includegraphics[scale=.46]{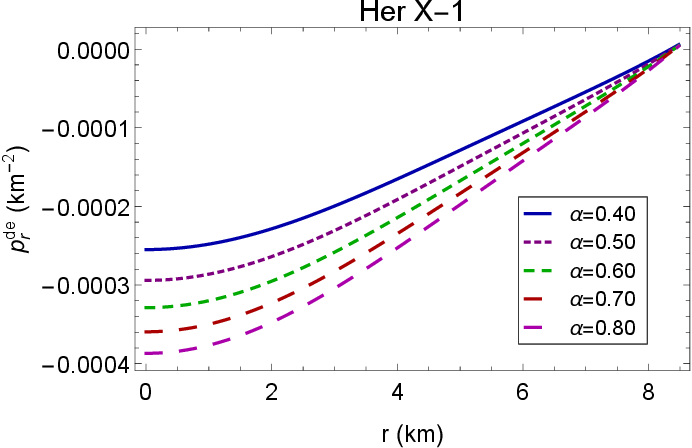}
         \includegraphics[scale=.46]{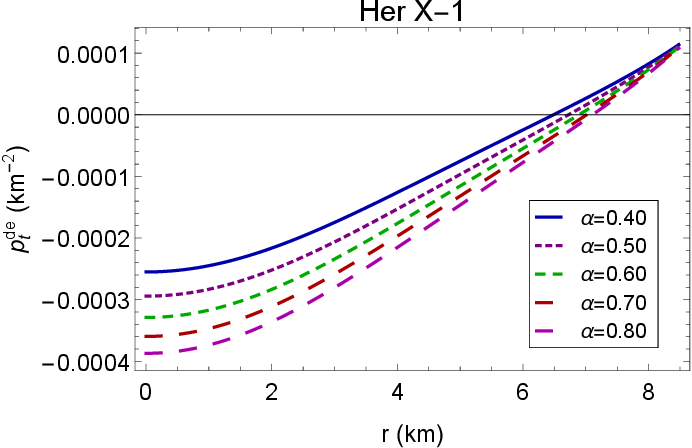}
        \caption{The variation of dark energy density and dark pressure with respect to `r'.}\label{dark}
\end{figure}

 
In this work, the repulsive nature of dark energy is simulated so that energy density remains positive ($\rho^{de}>0$) and radial dark pressure are negative ($p_r^{de}<0$). We have graphically displayed the profiles of dark energy density, radial dark pressure, and transverse dark pressure in Fig.~\ref{dark}. As derived from seed assumption ~(\ref{as3}), we can check from Fig.~\ref{dark} that transverse dark pressure $p_t^{de}$ remains negative in most of the regions and positive near stellar surface.
It is seen that dark energy density gradually decreases within the stellar configuration and its maximum value is at the center. Also from equation (\ref{de1}), we get the maximum value, $\rho^{de}_{max} = [\rho^{de}]_{r=0} = \frac{3a\alpha}{\kappa(1+\alpha)}$. Whereas, dark pressures are gradually increasing with $'r'$ whatever the value of $\alpha$ is. For our model, it is also interesting that at the center $(r = 0)$,
$p^{de}_r = p^{de}_t = - \rho^{de}_{max}$.

\subsection{Mass-radius relationship and compactness factor}

We know that the active mass depends on the density profile and it increases with the confining radius \cite{buchdahl1959general, glendenning2012compact} as it being gravitationally confined to a finite spatial extent $(r = R)$.
We can easily measure the mass function $m(r)$ for an electrically charged fluid sphere by computing the integral connected directly to the energy density (\ref{f11}) using the following expression \cite{florides1983complete, kumar2022isotropic},
\begin{eqnarray}
 m(r)=4\pi\int^r_0{\Big(\rho r^2+ r \sigma q e^{\lambda/2}\Big)\,dr},~~~\label{mm}
 \end{eqnarray}
 Now employing the metric potentials on (\ref{mm}) we obtain, 
\begin{eqnarray}\label{mm1} 
    m(r)=\frac{r}{2}\Big(1-e^{-\lambda(r)}+\frac{q^2}{r^2}\Big).
\end{eqnarray}

\begin{figure}[H]
    \centering
        \includegraphics[scale=.5]{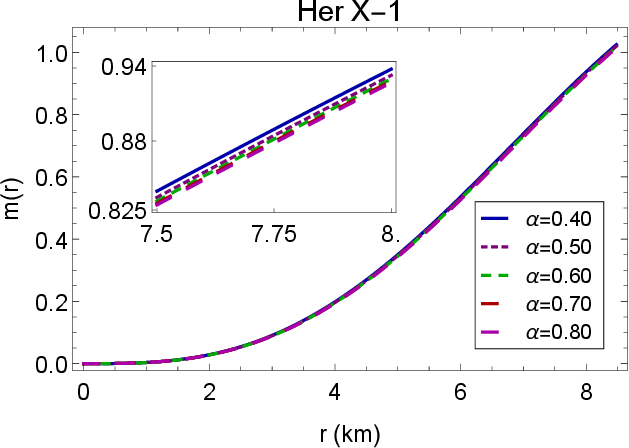}
        \includegraphics[scale=.5]{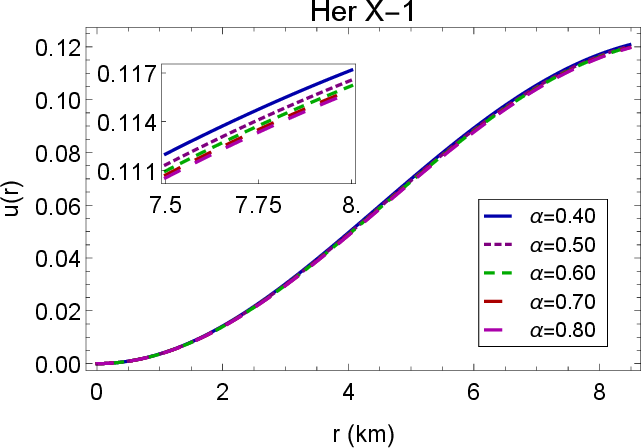}
       \caption{The mass function and compactness factor are plotted against `r' with required magnified inset.}\label{mass}
\end{figure}

This is to be noted that mass function $m(r)$ is a function of $r$ satisfying $m(r = 0) = 0$ whereas $m(r = R) = M$. The variation of mass function (\ref{mm}) has been plotted against $r$ with varying $\alpha$ in Fig.~\ref{mass}.\par
Clearly mass is regular at the center as it is directly proportional to the radial distance $r$ and maximum mass is attained at the surface $r=R$ as displayed in Fig.~\ref{mass}.\par
Furthermore, the compactness factor is determined by a dimensionless parameter $u(r) = \frac{m(r)}{r}$. The behavior of compactness factor $u(r)$ is shown in Fig.~\ref{mass} and see that $u(r)$ increases with $r$ for different values of $\alpha$.

\subsection{Surface redshift and Gravitational redshift}

{\bf The surface redshift $z_s(r)$ for our model is obtained by using the expression of compactness factor $u(r)$ as, $z_s(r)=\frac{1}{\sqrt{1-2u(r)}}-1$. We have shown the variation of $z_s(r)$ in Fig.~(\ref{red}) from center to surface. The surface redshift depends on the stellar mass and star radius, in other words on the surface gravity. 
In Table~\ref{tab1}, we present the numerical results of compactness factor ($u(R)$), surface redshift ($z_s(R)$) and mass (in $M_{\odot})$) for different values of the dark energy coupling parameter $\alpha$. From Table~\ref{tab1}, we notice that compactness factor ($u(R)$), surface redshift ($z_s(R)$) and mass (in $M_{\odot})$) decrease with increasing $\alpha$ in our chosen range $[0.40,~0.80]$.

\begin{figure}[H]
    \centering
        \includegraphics[scale=.52]{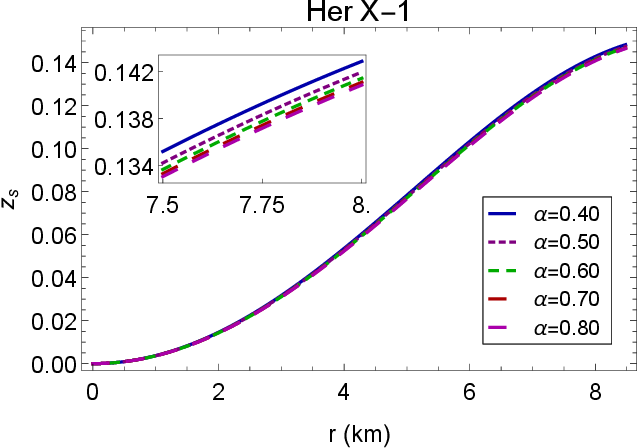}
        \includegraphics[scale=.46]{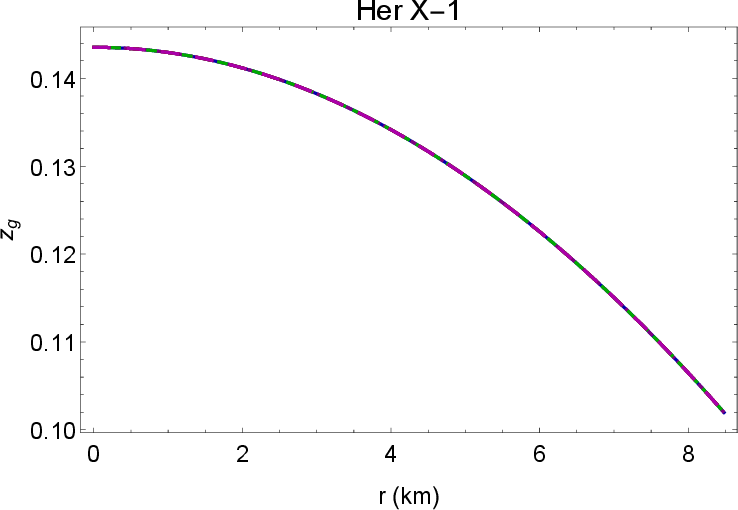}
         \caption{Surface redshift and gravitational redshift are plotted against `r' with required magnified inset.}\label{red}
\end{figure}
On the other hand, the gravitational redshift (or interior redshift)  $z_g(r)$ within a static line element can be expressed as,
$z_g(r)=e^{-\frac{\nu(r)}{2}}-1$. We see that  $z_g(r)$ is maximum at the center and gradually decreases with the increasing radius to attain its minimum at the surface.
We have plotted the variation of $z_g(r)$ in Fig.~(\ref{red}). 
From these figures, we can easily check that the surface redshift and gravitational redshift have totally opposite trends throughout our model.\par 
The actual reason behind this above trend is that if a photon travels from the center to the stellar surface, it must traverse a much denser region and a longer path, resulting in greater dispersion and a significant loss of energy. This energy must be lost by changing frequency rather than changing speed. When the energy of a photon decreases, so does its frequency. This results in a transition to the red end of the electromagnetic spectrum or an increase in the photon's wavelength. On another side, when a photon discharges from near the surface, it must travel through a less dense zone and along a shorter trajectory, resulting in lesser dispersion and less loss of energy. As a result, the gravitational (or internal) redshift is minimum at the surface and maximum at the center.
On the other side, the surface redshift is maximum near the surface and diminishes towards the core as the radius increases slightly with an increase in mass producing a larger surface gravity. Also from both figures, we can easily check that both types of redshifts have no singularity throughout their configuration.}

\begin{table}
\begin{center}
\caption{\label{tab1} The numerical values of $\alpha$, central density $(\rho_c)$, surface density $(\rho_s)$, central pressure $(p_c)$, compactness ratio $(u)$, surface redshift $(z_s)$, $p_c/\rho_c$, mass ($M_{\odot}$) have been shown for compact star, Her X-1.}
\begin{tabular}{cccccccc}
\hline
\hline
          & \multicolumn{7}{c}{Her X-1}  \\
\cline{1-8}
$\alpha$  & $\rho_c$  &   $\rho_s$  &   $p_c$  &       $u(R)$      &    $z_s(R)$  &   $p_c/\rho_c$ & $M (~M_{\odot})$ \\
  &  $\text{gm}~\text{cm}^{-3}$  &   $\text{gm}~\text{cm}^{-3}$  &$\text{dyne}~\text{cm}^{-2}$  &     &     &    &  \\  
\hline
     0.40 & $8.59982 \times 10^{14}$  & $4.63719 \times 10^{12}$  & $5.31679 \times 10^{13}$ & 0.120945 & 0.148508 & 0.0618245  & 1.02804 \\

     0.50 & $7.93748 \times 10^{14}$  & $4.29765 \times 10^{12}$  & $1.10500 \times 10^{14}$ & 0.120400 & 0.147682 & 0.139213   & 1.02340\\

     0.60 & $7.39378 \times 10^{14}$  & $4.00443 \times 10^{12}$  & $1.59792 \times 10^{14}$ & 0.120088 & 0.147212 & 0.216117   & 1.02075\\

     0.70 & $6.92988 \times 10^{14}$  & $3.74867 \times 10^{12}$  & $2.02898 \times 10^{14}$ & 0.119887 & 0.146908 & 0.292788   & 1.01904\\

     0.80 & $6.52575 \times 10^{14}$  & $3.52362 \times 10^{12}$  & $2.41015 \times 10^{14}$ & 0.119746 & 0.146695 & 0.369329   & 1.01784\\

\hline
\hline
\end{tabular}
\end{center}
\end{table}

\section{Stability analysis of our model and equilibrium of forces}\label{stable}

\subsection{Critical stability bound on Mass-radius ratio}
The mass-radius (M-R) ratio of a compact star cannot be arbitrarily large for physically realistic models. According to Buchdahl \cite{buchdahl1959general}, for a perfect fluid sphere model without charge, the M-R ratio must satisfy the inequality $\frac{2M}{R}<\frac{8}{9}$. This upper limit of the M-R ratio was computed using a perfect fluid with decreasing energy density near the boundary. But the presence of electric charge modifies this Buchdahl limit. B\"Ohmer and Harko \cite{Boehmer:2007gq} suggested the generalized expression of the lower limit for a charged compact object as given following:
\begin{eqnarray} \label{bd1}
\frac{Q^{2}\left(18\,R^{2}+Q^{2}\right)}{R^{2}\left(12\,R^{2}+Q^{2}\right)} \leq \frac{2M}{R}.
\end{eqnarray}
Andr\'easson \cite{Andreasson:2008xw} then showed that, for a relativistic charged sphere, to attain critical stability the model must satisfy the following inequality:

\begin{eqnarray} \label{bd2}
  \sqrt{M} \leq   \frac{\sqrt{R}}{3} + \sqrt{\frac{R}{9}+\frac{Q^2}{3R}}.
\end{eqnarray}

Now combining these two inequalities (\ref{bd1}) and (\ref{bd2}), we get both lower and upper bound of $\frac{M}{R}$ as given following:

\begin{eqnarray} \label{bd3}
\frac{Q^{2}\left(18\,R^{2}+Q^{2}\right)}{2\,R^{2}\left(12\,R^{2}+Q^{2}\right)} \leq \frac{{M}}{R}\leq \frac{1}{R}\Bigg[\frac{\sqrt{R}}{3} + \sqrt{\frac{R}{9}+\frac{Q^2}{3R}}\Bigg]^2.
\end{eqnarray}
The above inequality (\ref{bd3}) is sharp and this sharpness is attained by arbitrarily thin shell solutions \cite{Andreasson:2008xw}.

\subsection{Causality condition via Herrera's cracking approach}

\begin{figure}[H]
    \centering
        \includegraphics[scale=.47]{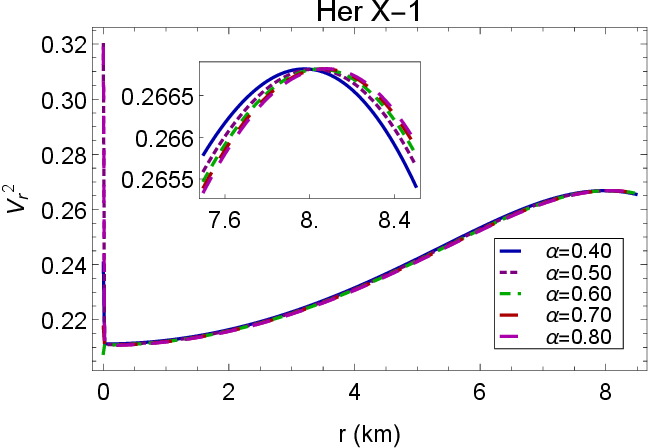}
        \includegraphics[scale=.47]{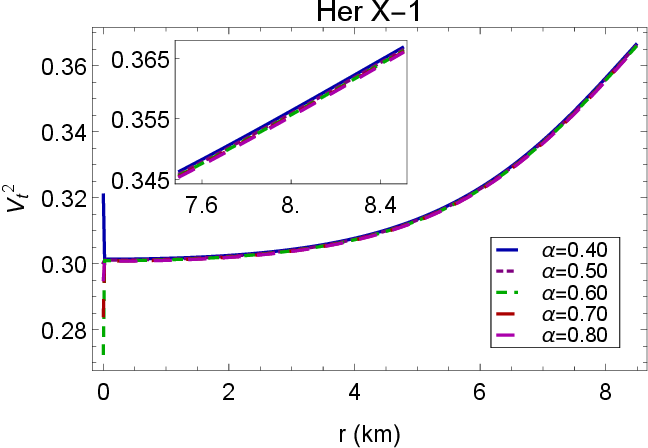}
        \includegraphics[scale=.47]{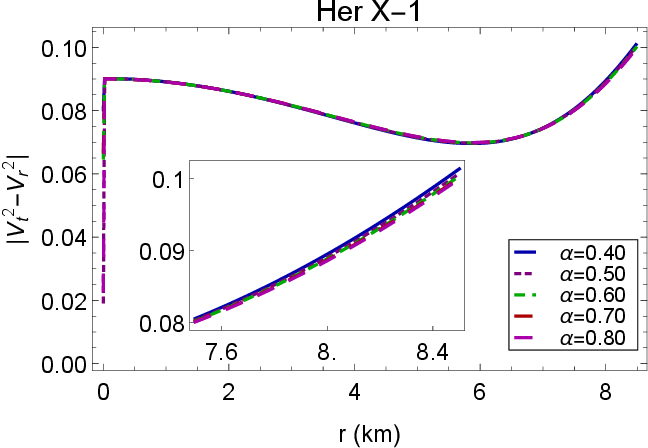}
        \caption{Square of sound velocity components and the stability factor $|V_t^2-V_r^2|$ are plotted against `r' with required magnified inset.}\label{sv}
\end{figure}

\textbf{Now in this subsection, we will discuss another important "physical acceptability condition" for realistic models known as the causality condition using sound velocity along with Herrera's cracking approach. We start by talking about our model's causality condition, which states that the square of sound velocity $V^2$=$\frac{dp}{d\rho}$ should be less than unity for a physically realistic model \cite{Herrera:1992lwz, Abreu:2007ew}. This means that the speed of sound is not faster than the speed of light. Thus employing expressions~(\ref{ef1})-(\ref{ef3}), we calculate the effective radial and tangential velocity components for our anisotropic model as,
\begin{eqnarray}
V_r^2&=&\frac{dp_r^{\text{eff}}}{d\rho^{\text{eff}}} ~~\rm{and} \label{sp1}\\
V_t^2&=&\frac{dp_t^{\text{eff}}}{d\rho^{\text{eff}}}.\label{sp2}
\end{eqnarray}
Because of the complicacy of their analytic expressions, we have analyzed them graphically in Fig.~(\ref{sv}) and observe that both lie between the desired range $(0, 1)$ within the stellar object. We clearly notice that the sound velocity components are always positive irrespective of matter density. Hence the causality condition is satisfied for our proposed charged dark energy star model. The unnatural behavior at the initial of the plots of sound velocity components is due to the consideration of well-known dark energy EoS (\ref{as2}) which is similar to the MIT Bag model for hadrons, which may define the quark matter inside the stellar system. Thus these kinds of behavior occur in these plots due to the presence of quark matter inside the stellar system.\\
Moreover, Herrera has put out the "cracking" (or overturning) technique \cite{Herrera:1992lwz} for relativistic stellar objects under minor radial perturbations. Abreu et al. used the idea of cracking in their research \cite{Abreu:2007ew} and propose the idea of stability factor, mathematically defined as $|V_t^2-V_r^2|<1$. The profile of this stability factor is also plotted in Fig.~(\ref{sv}) and we see that this condition is also satisfied throughout our model. Henceforth, our model is physically stable and well-behaved.}

\subsection{Energy conditions}

\begin{figure}[H]
    \centering
        \includegraphics[scale=.47]{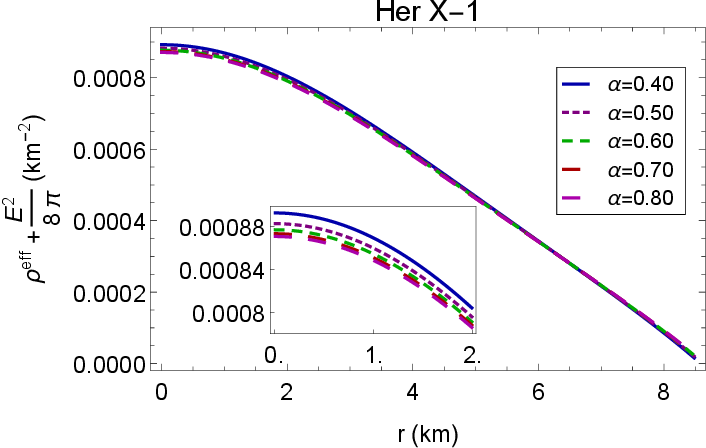}
        \includegraphics[scale=.47]{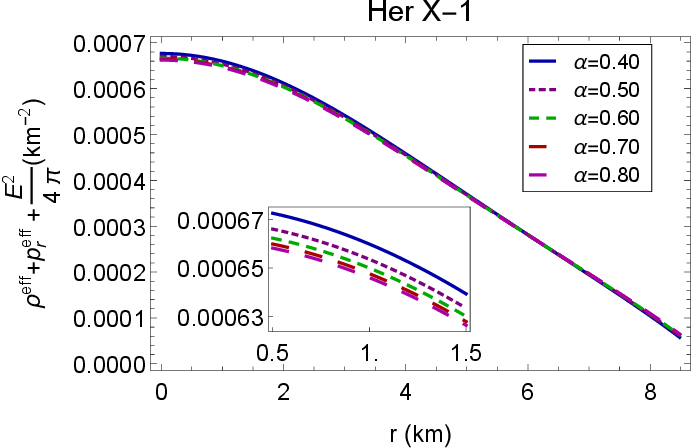}
        \includegraphics[scale=.47]{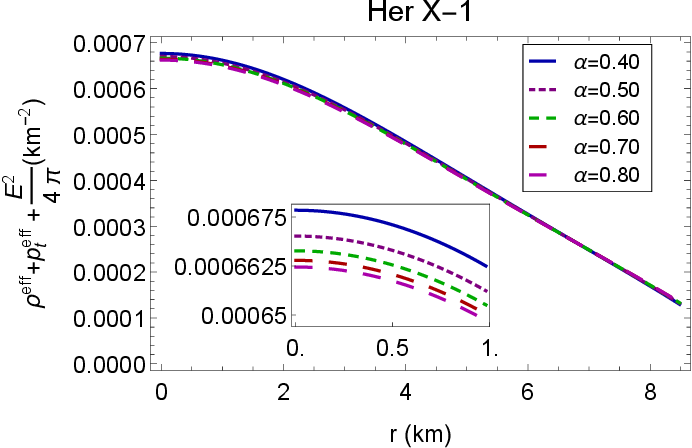}
        \includegraphics[scale=.47]{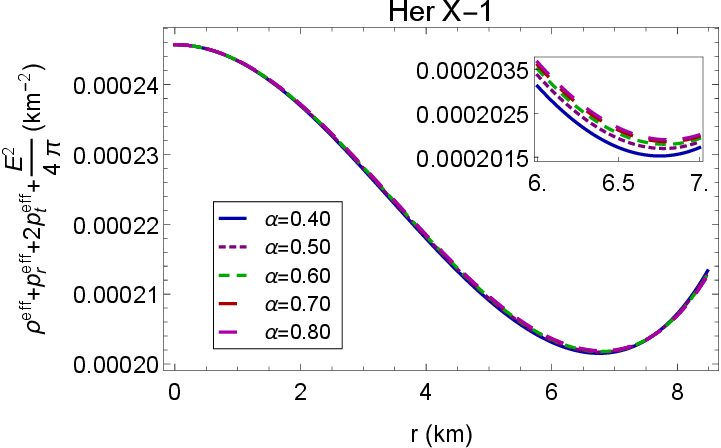}
        \includegraphics[scale=.47]{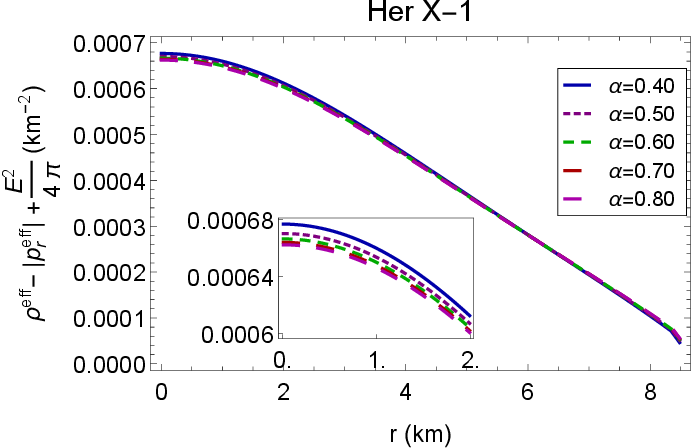}
        \includegraphics[scale=.47]{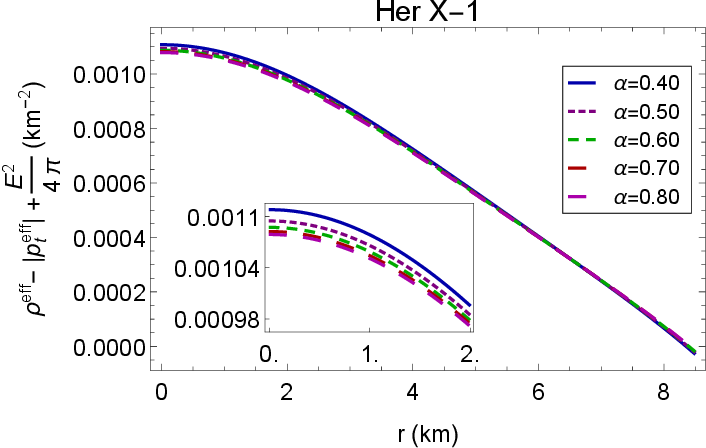}
         \caption{Behaviour of energy conditions vs. radius `r' with required magnified inset.}\label{ec}
\end{figure}

{\bf Here we shall investigate the energy conditions (ECs) of a charged fluid sphere model coupled to anisotropic dark energy according to the relativistic classical field theories of gravitation. In the context of GR, the ECs are local
inequalities that define a relation between matter-energy density and pressure that obeys certain bounds. An anisotropic-charged fluid sphere should satisfy the four standard, model-independent, pointwise ECs termed as (i) Null energy condition (NEC), (ii) Weak energy condition (WEC), (iii) Strong energy condition (SEC), and (iv) Dominant energy condition (DEC) \cite{bondi1947spherically, witten1981new, visser1997energy, Andreasson:2008xw, garcia2011energy}. For our present model, the corresponding effective energy conditions must hold simultaneously throughout the stellar model and these are given by these following inequalities :

\begin{itemize}
\item NEC:~$\rho^{\text{eff}}+ p_r^{\text{eff}} + \frac{E^2}{4\pi} \geq 0,~\rho^{\text{eff}}+ p_t^{\text{eff}} + \frac{E^2}{4\pi} \geq 0$, 
\item WEC:~$\rho^{\text{eff}}+ p_r^{\text{eff}} + \frac{E^2}{4\pi} \geq 0,~\rho^{\text{eff}}+ p_t^{\text{eff}} + \frac{E^2}{4\pi} \geq 0,~ \rho^{\text{eff}} + \frac{E^2}{8\pi}  \geq 0$, 
\item SEC:~$\rho^{\text{eff}}+ p_t^{\text{eff}} + \frac{E^2}{4\pi} \geq 0,~\rho^{\text{eff}} + p_r^{\text{eff}} + 2p_t^{\text{eff}}+ \frac{E^2}{4\pi} \geq 0$, 
\item DEC:~$\rho^{\text{eff}}-|p_r^{\text{eff}}| + \frac{E^2}{4\pi} \geq 0,~ \rho^{\text{eff}}-|p_t^{\text{eff}}| + \frac{E^2}{4\pi} \geq 0,~\rho^{\text{eff}} + \frac{E^2}{8\pi} \geq 0$.
\end{itemize}
From Fig.~(\ref{ec}), we clearly see that our proposed dark energy star model satisfies the above bounds of all energy conditions straightforwardly for all $r$. Hence, our model is physically realistic.}

\subsection{Hydrostatic Equilibrium via Modified TOV Equation}
Now we shall investigate the hydrostatic equilibrium for our present model in the presence of dark energy. To check the equilibrium of our model under various forces acting on it, we shall employ the Tolman-Oppenheimer-Volkoff equation, which is described by \cite{Ghezzi:2009ct, ponce1993limiting}:
\begin{equation}\label{tov1}
-\frac{M_G(r)(\rho+p)}{r}e^{(\lambda-\nu)/2}-\frac{dp}{dr}-\frac{dp_r^{de}}{dr}+\frac{2}{r}(p_t^{de}-p_r^{de})+\sigma Ee^{\lambda/2}=0,
\end{equation}
proposed by Tolman, Oppenheimer, and Volkoff and thus named as {\em TOV} equation.\\
Where $M_G(r)$ is effective gravitational mass inside a sphere of radius $r$  as obtained from the Tolman-Whittaker formula \cite{PhysRev.35.875} and Einstein's field equations and is expressed as,
\begin{equation}\label{tov2}
M_G(r)=\frac{1}{2}r^2 \nu' e^{(\nu - \lambda)/2}.
\end{equation}

Substituting the expression of $M_G(r)$ in equation \eqref{tov1}, we obtain,
\begin{equation}\label{tov1a}
-\frac{\nu'}{2}(\rho+p)-\frac{dp}{dr}-\frac{dp_r^{de}}{dr}+\frac{2}{r}(p_t^{de}-p_r^{de})+\sigma Ee^{\lambda/2}=0.
\end{equation}

The above equation (\ref{tov2}) can also be written as,
\begin{equation}
F_g + F_h + F_d + F_e=0,
\end{equation}
where $F_g, F_h$, $F_e$ represent the gravitational, hydrostatic-gradient, and electric force respectively whereas  another force term $F_d$ arises due to dark energy given as follows:
\begin{eqnarray}
\text{Gravitational force:}~ F_g&=&-\frac{\nu'}{2}(\rho+p) \\
\text{Hydrostatic-gradient force:}~ F_h&=& -\frac{dp}{dr}\\
\text{Dark energy force:}~ F_d&=& -\frac{dp_r^{de}}{dr}+\frac{2}{r}(p_t^{de}-p_r^{de})\\
\text{Electric force:}~ F_e&=& \sigma Ee^{\lambda/2}.
\end{eqnarray}

\begin{figure}[H]
    \centering
        \includegraphics[scale=.465]{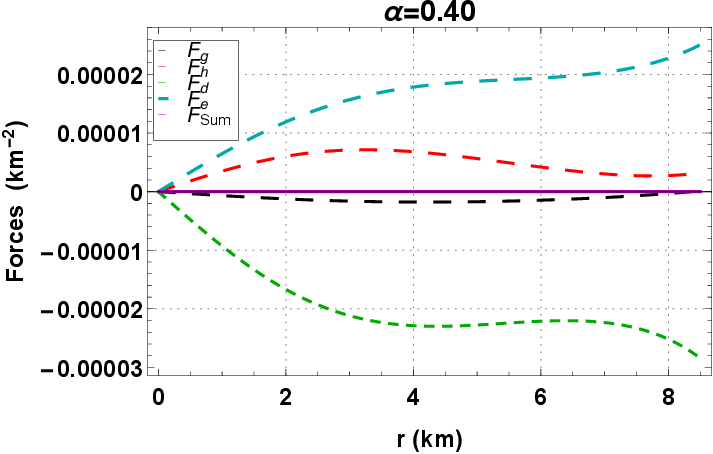}
        \includegraphics[scale=.465]{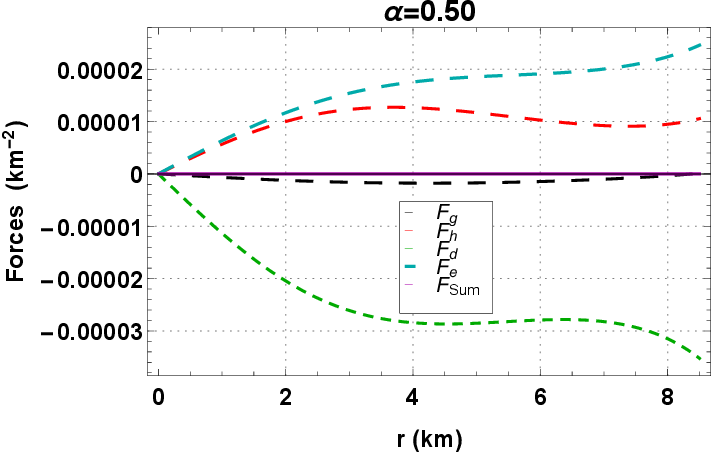}
        \includegraphics[scale=.465]{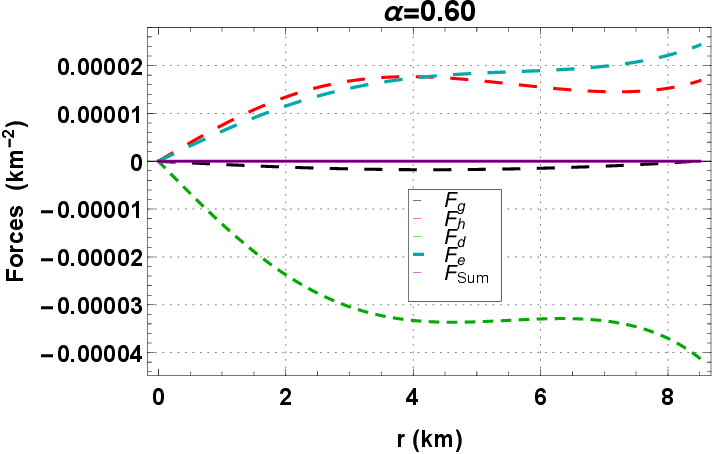}
        \includegraphics[scale=.465]{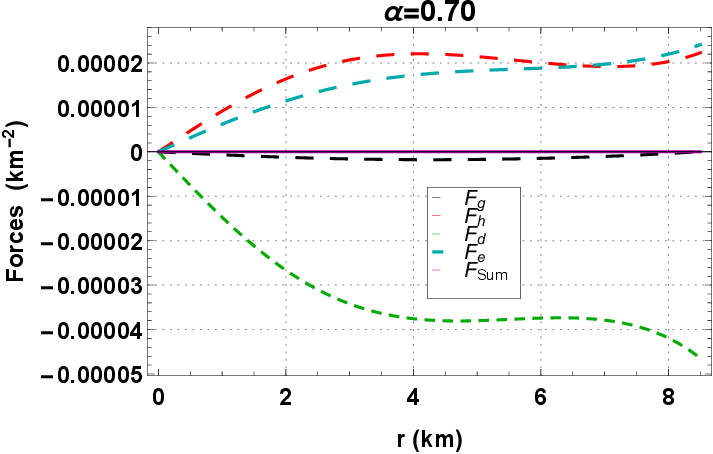}
        \includegraphics[scale=.465]{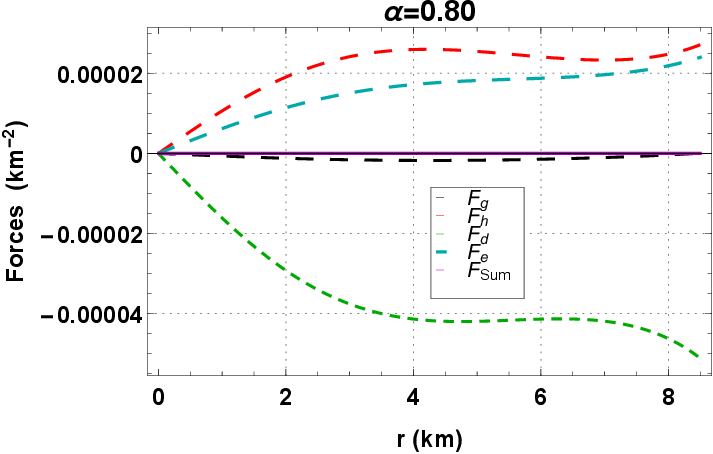}
        \caption{Behavior of different forces, such as the gravitational force $F_g$, hydrostatic gradient force $F_h$, dark energy force $F_d$ and electric force $F_e$ respectively versus radial coordinate $r$ for different values of $\alpha$. Here $F_{\text{Sum}}=F_g + F_h +F_d + F_e$.}
    \label{tov}
\end{figure}
We can check the behavior of the generalized TOV equations from Fig.~\ref{tov} and the system is counterbalanced by all these four forces present in our system. This concludes that our model attains a static equilibrium. In this context, it is also noted that the hydrostatic gradient ($F_h$) and electric forces ($F_e$) are positive, whereas the gravitational ($F_g$) and dark energy forces ($F_d$) are negative. Also to properly justify the counterbalance effect of all these four forces, we have also plotted the overall effect of all four forces $F_{\text{Sum}}=F_g + F_h +F_d + F_e$ and see that it vanishes (Magenta line of Fig.~\ref{tov}). This concludes that our present model attains a static equilibrium.

\subsection{Equation of state via Zel'dovich condition}
The Zel'dovich requirement \cite{shapiro2008black, zeldovich1971relativistic, zel1972relativistic, l1962equation} for the stability of a stellar structure is that the ratio of pressure to density must be less than unity everywhere within the stellar interior. If we define the ratios as: $\Omega(r) = \frac{p(r)}{\rho(r)}$, then $\Omega(r)<1$.  

\begin{figure}[H]
    \centering
        \includegraphics[scale=.52]{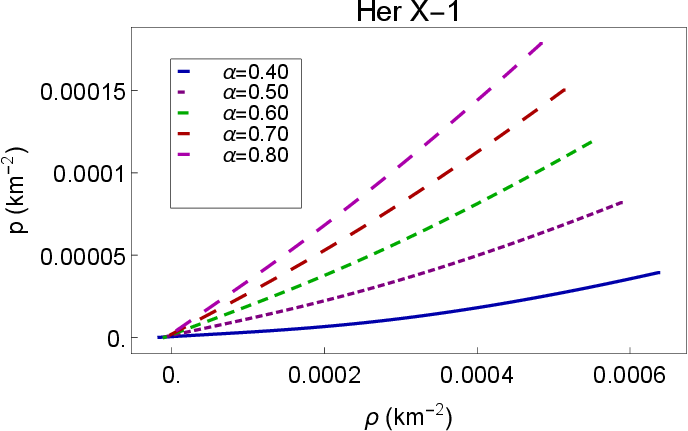}
        \includegraphics[scale=.5]{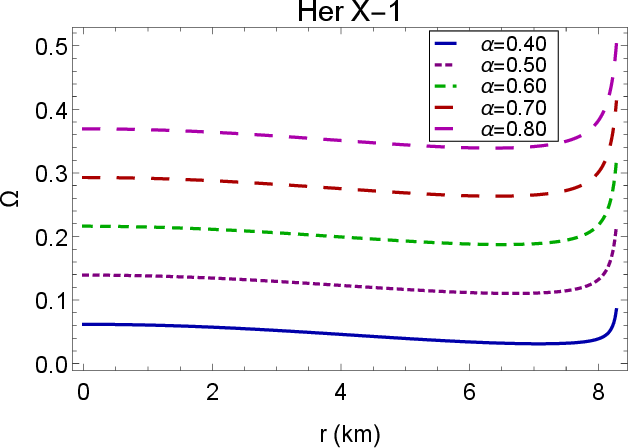}
         \includegraphics[scale=.35]{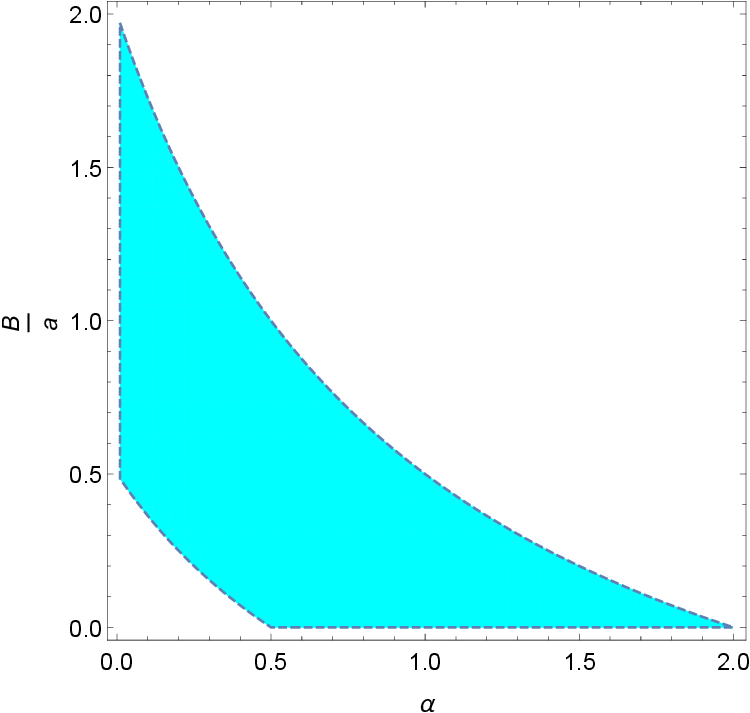}
        \caption{(i)Variation of pressure $p$ with respect to density $\rho$, (ii) the ratio $\Omega$ are shown inside the stellar interior, and (iii) constraint region plot of $B/a$ with respect to $\alpha$.}\label{eos}
\end{figure}

From Fig.~(\ref{eos}), it is clear that our model satisfies the Zel'dovich ratio $\Omega(r)<1$ for all $r < R$ and for all dark energy coupling parameters, $\alpha \in [ 0.40, 0.80 ]$. This $\Omega$ explains the idea of the equation of state parameters and it is a key factor to determine the stellar formations. \par
Now, if we apply the Zel'dovich condition for the central values of stellar parameters $\rho$ ~and~ $p$, we get $\frac{p_c}{\rho_c} < 1$, from which we obtain the relationship,
\begin{eqnarray}
\frac{a (-1 + 2 \alpha) + 2 (1 + \alpha) B}{3 a} < 1 \label{es1}
\end{eqnarray}
which implies that,
\begin{eqnarray}
\frac{B}{a}  < \frac{2-\alpha}{1+\alpha} \label{es2}
\end{eqnarray}
Also from the relation $p_c > 0$ we get,
\begin{eqnarray}
\frac{B}{a}  > \frac{1- 2\alpha}{2(1+\alpha)} \label{es3}
\end{eqnarray}
Now combining relations (\ref{es2}) and (\ref{es3}), we finally obtain a boundary representing constraint on $\frac{B}{a}$ in the following form: 
\begin{eqnarray}
 \frac{1- 2\alpha}{2(1+\alpha)} < \frac{B}{a} < \frac{2-\alpha}{1+\alpha} \label{es4}
\end{eqnarray}
In Fig.~(\ref{eos}), we have portrayed the constraint region for $B/a$ given in the expression (\ref{es4}) with respect to $\alpha$.
    
\subsection{Harrison-Zel'dovich-Novikov's static stability criterion}

Harrison-Zel'dovich-Novikov \cite{zeldovich1971relativistic, harrison1965gravitation} proposed the criterion for stability of any compact object. According to their criterion, any fluid configuration is stable if its total mass is increasing with increasing central density, i.e. $\frac{\partial M (\rho_c)}{\partial \rho_c} > 0$, whereas is unstable if the total mass of the star is decreasing with respect to the central density i.e. $\frac{\partial M (\rho_c)}{\partial \rho_c} < 0$. For this purpose, we have to calculate the mass ($M$) and its gradient in terms of central density as,
\begin{eqnarray}
\rho_c &=& \frac {3 a}{\kappa (1 +  \alpha)}
\end{eqnarray}
For our present model,
\begin{eqnarray}\label{har1}
M (\rho_c) &=& \frac{1}{8 \Big(3 + 3 b R^4 + (1 + \alpha) \kappa \rho_c R^2 \Big)^2}\Bigg[9 R^5 (B^2 + b \Big(3 - 2 B R^2 + (5 b + B^2) R^4)\Big) + 
 3 (1 + \alpha) \kappa R^3 \Big(4 - B R^2 \nonumber\\ &&  
+(10 b + B^2) R^4\Big) \rho_c + 5 (1 + \alpha)^2 \kappa^2 R^5 \rho_c^2\Bigg]
\end{eqnarray}
Now taking partial derivative with respect to $\rho_c$ we get,
\begin{eqnarray}\label{har2}
\frac{\partial M (\rho_c)}{\partial \rho_c} &=& \frac{1}{8 \Big(3 + 3 b R^4 + (1 + \alpha) \kappa \rho_c R^2 \Big)^3}\Bigg[3\kappa R^3 (1 + \alpha) \Bigg\{12 + R^2 \Bigg(-3 \Big(B - 8 b R^2 - 3 b B R^4 + B^2 (R^2 + b R^6)\Big)  \nonumber\\ &&  - \kappa \rho_c(1 +\alpha)  (-3 + B R^2) (2 + B R^2)  \Bigg)\Bigg\}\Bigg]
\end{eqnarray}
\begin{figure}[H]
    \centering
        \includegraphics[scale=.55]{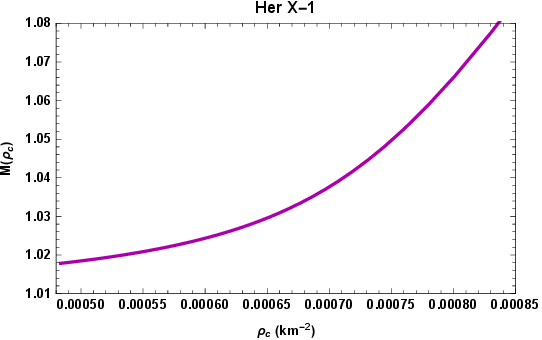}
        \includegraphics[scale=.55]{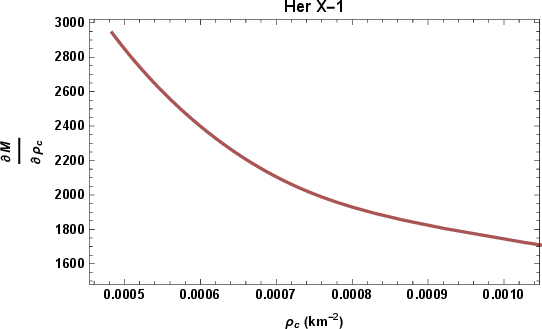}
         \caption{Behavior of Mass ($M (\rho_c)$) and its gradient $\frac{\partial M (\rho_c)}{\partial \rho_c}$ with respect to central density $\rho_c$ for the compact star Her X-1.}\label{har}
\end{figure}
From Fig.~(\ref{har}), we can easily verify that the mass ($M (\rho_c)$) increases with central density $\rho_c$. On the other hand, we notice that the gradient $\frac{\partial M (\rho_c)}{\partial \rho_c}$ becomes monotonically decreasing with increasing central density $\rho_c$ although it remains positive with respect to central density $\rho_c$ throughout the stellar interior. This implies that our current stellar model is stable under Harrison-Zel'dovich-Novikov stability criterion.

\section{Discussion and Concluding Remarks}\label{con}

In this present article, we have made an attempt to model a unique charged strange star model coupled with inhomogeneous anisotropic dark energy. We have taken into account the well-known Tolman-Kuchowicz {\em ansatz} to solve the Einstein-Maxwell field equations for a static symmetric perfect fluid distribution. We also choose  a particular form of EoS. The dark energy follows the distribution of matter at high energy density, with a strength that depends on the parameter $\alpha$. The fundamental distinction of this type of investigation is that in the original dark energy stellar model, the core is de-Sitter and the shell is close to or replaces the event horizon. Furthermore, there exists a critical parameter $\alpha_c$, that distinguishes dark energy stars from conventional black hole solutions.

 Several research works have been done earlier on dark energy stars composed of ordinary fluid and dark energy. According to Shvartsman \cite{1971JETP...33..475S}, stars can carry electric charges. So here we present a model of singularity free charged dark energy star. For many reasons, including its significance as a possible substitute for a black hole, the charged dark energy star model has gained a bigger astrophysical relevance. Also, it has been found that dark energy along with anisotropic stress can help to produce stability for the stars against gravitational collapse as it exerts a repulsive force looking similar to electric charge, on dark energy surrounding.\par
Here we develop a straightforward numerical technique to integrate the stellar structure equations from the core of the star to its boundary. The compactness of the stars, their structural parameters, and M-R relations have been calculated to compare with the well-known theoretical TOV model.

 For this theoretical investigation, we have chosen a particular compact dark energy star HER X-1 to analyze our results numerically as well as graphically throughout this paper. We plot all physical properties for dark energy coupling parameter values $\alpha = 0.40, 0.50, 0.60, 0.70, 0.80$. Here, to determine the constants present in the TK metric coefficients, we smoothly match the interior metric with the exterior Reissner-Nordstr\"{o}m metric by assuming continuity of the metric functions $g_{tt}$, $g_{rr}$ and $\frac{\partial g_{tt}}{\partial r}$ across the boundary surface $r= R$ along with $p(r=R)=0$.
We require to match our results with available observational data. Because the model meets all of the necessary physical conditions and is horizon free, so this type of star may be realistic in our nature. 

The following are the salient key features of our proposed dark energy stellar model:
\begin{enumerate}
  \item The graphical representations of $e^{\nu(r)}$ and $e^{-\lambda(r)}$ in Fig.~(\ref{metric}) clearly indicate that they are finite and nonsingular over the radius of stars for varying $\alpha$, which is necessary to generate physically viable models of compact stellar objects.
  
   \item It is clear from Fig.~(\ref{rho}) that the energy density $\rho(r)$ and pressure $p(r)$ remain continuous, positively finite inside the star, and demonstrate a smooth declining nature towards the surface for various values of $\alpha$. Pressure and density vanish at the boundary $r = R$. We see that $\rho$ and $p$ are maximum at the core ($r = 0$). Both density and pressure are at their highest levels at the core, which indicates the presence of highly compact cores. Whereas electric field intensity $E^2$ increases with $r$. It remains positive and monotonic increasing throughout the fluid sphere.
   \item The gradient components of matter-energy density and pressure show a negative trend in Fig.~(\ref{grad}) i.e. from zero to negative region whenever $r$ goes from center to boundary. This scenario remains the same for different considerations of $\alpha$. The pressure and density gradients vanish at the center ($r=0$) and remain negative inside the stellar configuration. Also, double derivatives take negative values at the center which implies that both density and pressure take maximum value at the center of the stellar object.

   \item This model represents the repulsive character of dark energy, which leads to the fact that radial dark pressure $(p^{de}_r)$ is negative while energy density $(\rho^{de})$ is positive (From Fig.~\ref{dark}) whereas in most of the stellar regions, transverse dark pressure $(p^{de}_t)$ is still negative and positive near the stellar surface.

    \item In Table~\ref{tab1}, the numerical values of central density $(\rho_c)$, surface density $(\rho_s)$, central pressure $(p_c)$, compactness ratio $(u(R))$, surface redshift $(z_s(R))$, $p_c/\rho_c$, and mass ($M_{\odot}$) have been presented for different values of $\alpha$.
    From Fig.~(\ref{mass}), it is obvious that the mass function and compactness factor are regular throughout the stellar region and gradually increase with 'r'. Also from Fig.~(\ref{red}), we notice the interesting fact that the gravitational (or internal) redshift $(z_g)$ is minimum at the surface and maximum at the center, in contrast, the surface redshift $(z_s)$ behaves just opposite to $z_g$. 
  
    \item Our proposed charged dark energy stellar model also satisfies the Causality condition i.e. square of the sound velocity $V^2$ lies within $(0,1)$ (From Fig.~\ref{sv}) within the stellar body which leads the model to be physically stable and well-behaved.

  \item Our obtained model satisfies all four energy conditions viz., NEC, WEC, SEC, and DEC for various values of $\alpha$, and that is illustrated graphically in Fig.~(\ref{ec}). Moreover, these always remain positive in the entire region of the star and it confirms the viability of our proposed solution.

  \item  Full illustration for the profiles of $F_g$, $F_h$, $F_d$, and $F_e$ for our proposed stellar structure have been made in Fig.~(\ref{tov}). By combining the effects of all these forces, it is obviously conceivable to create a static equilibrium configuration. Here, hydrostatic-gradient and electric forces are repulsive whereas gravitational force and dark energy force are attractive in nature.
  
 \item In Fig.~(\ref{eos}), we analyze the equation of state(EoS) parameters via Zel'dovich condition for stability i.e $\Omega(r)=\frac{p(r)}{\rho(r)}$ should be less than unity, is also verified even though the star comprises electric charge, ordinary matter, and dark energy. So it is obvious that the matter content is non-exotic in nature and the solution of our proposed model is physically well-behaved. Also, we plot $p$ with respect to $\rho$ and the constraint region for $B/a$ with respect to $\alpha$ in Fig.~(\ref{eos}).
  
\item From the inequality~(\ref{bd3}), we can reach the conclusion that the respective M-R ratio values for the presence of anisotropies and electric charge in the matter distribution are higher than the corresponding ones for the isotropic uncharged matter configurations.

\item Finally we analyze the stability situation through Harrison-Zel'dovich-Novikov's stability condition. We plot mass ($M (\rho_c)$) as a function of central density ($\rho_c$) and its gradient $\frac{\partial M (\rho_c)}{\partial \rho_c}$ in Fig.~(\ref{har}). Thus we arrive at a decision that our current stellar model is stable because it meets this Harrison-Zel'dovich-Novikov stability criterion.

\end{enumerate}

Therefore, with respect to all the significant results, we can finally conclude that we are able to structure a physically acceptable, stable, and singularity-free generalized model for charged strange stars employing the dark energy equation of state, which is suitable for studying dark energy stars within the Einstein gravity framework. We expect that our model may hypothetically contribute to the astrophysical scenario on a larger scale.

\section*{Author contributions}
\textbf{Pramit Rej} contributed to the conceptualization, editing, mathematical analysis, designing of computer codes for data analysis, revision works and overall supervision of the study. \textbf{Akashdip Karmakar} performed mathematical analysis, running computer codes, validation, and typing of the manuscript. All authors read and approved the final manuscript.

\section*{Declarations}
\textbf{Funding:} The authors did not receive any funding in the form of financial aid or grant from any institution or organization for the present research work.\par
\textbf{Data Availability Statement:} The results are obtained
via purely theoretical calculations and can be verified analytically,
thus this manuscript has no associated data, or the data will not be deposited. \par
\textbf{Conflicts of Interest:} The authors have no financial interest or involvement which is relevant by any means to the content of this study.

\bibliography{de_references}

\begin{thebibliography}{85}
\expandafter\ifx\csname natexlab\endcsname\relax\def\natexlab#1{#1}\fi
\expandafter\ifx\csname bibnamefont\endcsname\relax
  \def\bibnamefont#1{#1}\fi
\expandafter\ifx\csname bibfnamefont\endcsname\relax
  \def\bibfnamefont#1{#1}\fi
\expandafter\ifx\csname citenamefont\endcsname\relax
  \def\citenamefont#1{#1}\fi
\expandafter\ifx\csname url\endcsname\relax
  \def\url#1{\texttt{#1}}\fi
\expandafter\ifx\csname urlprefix\endcsname\relax\def\urlprefix{URL }\fi
\providecommand{\bibinfo}[2]{#2}
\providecommand{\eprint}[2][]{\url{#2}}

\bibitem[{\citenamefont{Maurya and Tello-Ortiz}(2019)}]{Maurya:2019zyc}
\bibinfo{author}{\bibfnamefont{S.~K.} \bibnamefont{Maurya}} \bibnamefont{and}
  \bibinfo{author}{\bibfnamefont{F.}~\bibnamefont{Tello-Ortiz}},
  \bibinfo{journal}{Eur. Phys. J. C} \textbf{\bibinfo{volume}{79}},
  \bibinfo{pages}{33} (\bibinfo{year}{2019}).

\bibitem[{\citenamefont{Tamta and Fuloria}(2017)}]{tamta2017new}
\bibinfo{author}{\bibfnamefont{R.}~\bibnamefont{Tamta}} \bibnamefont{and}
  \bibinfo{author}{\bibfnamefont{P.}~\bibnamefont{Fuloria}},
  \bibinfo{journal}{Journal of Modern Physics} \textbf{\bibinfo{volume}{8}},
  \bibinfo{pages}{1762} (\bibinfo{year}{2017}).

\bibitem[{\citenamefont{Bondi}(1992)}]{bondi1992anisotropic}
\bibinfo{author}{\bibfnamefont{H.}~\bibnamefont{Bondi}},
  \bibinfo{journal}{Monthly Notices of the Royal Astronomical Society}
  \textbf{\bibinfo{volume}{259}}, \bibinfo{pages}{365} (\bibinfo{year}{1992}).

\bibitem[{\citenamefont{Deb et~al.}(2017)\citenamefont{Deb, Chowdhury, Ray,
  Rahaman, and Guha}}]{DEB2017239}
\bibinfo{author}{\bibfnamefont{D.}~\bibnamefont{Deb}},
  \bibinfo{author}{\bibfnamefont{S.~R.} \bibnamefont{Chowdhury}},
  \bibinfo{author}{\bibfnamefont{S.}~\bibnamefont{Ray}},
  \bibinfo{author}{\bibfnamefont{F.}~\bibnamefont{Rahaman}}, \bibnamefont{and}
  \bibinfo{author}{\bibfnamefont{B.}~\bibnamefont{Guha}},
  \bibinfo{journal}{Annals of Physics} \textbf{\bibinfo{volume}{387}},
  \bibinfo{pages}{239} (\bibinfo{year}{2017}), ISSN \bibinfo{issn}{0003-4916},
  \urlprefix\url{https://www.sciencedirect.com/science/article/pii/S0003491617302920}.

\bibitem[{\citenamefont{Dev and Gleiser}(2002)}]{dev2002anisotropic}
\bibinfo{author}{\bibfnamefont{K.}~\bibnamefont{Dev}} \bibnamefont{and}
  \bibinfo{author}{\bibfnamefont{M.}~\bibnamefont{Gleiser}},
  \bibinfo{journal}{General relativity and gravitation}
  \textbf{\bibinfo{volume}{34}}, \bibinfo{pages}{1793} (\bibinfo{year}{2002}).

\bibitem[{\citenamefont{Di~Prisco et~al.}(1997)\citenamefont{Di~Prisco,
  Herrera, and Varela}}]{di1997cracking}
\bibinfo{author}{\bibfnamefont{A.}~\bibnamefont{Di~Prisco}},
  \bibinfo{author}{\bibfnamefont{L.}~\bibnamefont{Herrera}}, \bibnamefont{and}
  \bibinfo{author}{\bibfnamefont{V.}~\bibnamefont{Varela}},
  \bibinfo{journal}{General Relativity and Gravitation}
  \textbf{\bibinfo{volume}{29}}, \bibinfo{pages}{1239} (\bibinfo{year}{1997}).

\bibitem[{\citenamefont{Krori et~al.}(1984)\citenamefont{Krori, Borgohain, and
  Devi}}]{krori1984some}
\bibinfo{author}{\bibfnamefont{K.}~\bibnamefont{Krori}},
  \bibinfo{author}{\bibfnamefont{P.}~\bibnamefont{Borgohain}},
  \bibnamefont{and} \bibinfo{author}{\bibfnamefont{R.}~\bibnamefont{Devi}},
  \bibinfo{journal}{Canadian journal of physics} \textbf{\bibinfo{volume}{62}},
  \bibinfo{pages}{239} (\bibinfo{year}{1984}).

\bibitem[{\citenamefont{Mak and Harko}(2003)}]{mak2003anisotropic}
\bibinfo{author}{\bibfnamefont{M.}~\bibnamefont{Mak}} \bibnamefont{and}
  \bibinfo{author}{\bibfnamefont{T.}~\bibnamefont{Harko}},
  \bibinfo{journal}{Proceedings of the Royal Society of London. Series A:
  Mathematical, Physical and Engineering Sciences}
  \textbf{\bibinfo{volume}{459}}, \bibinfo{pages}{393} (\bibinfo{year}{2003}).

\bibitem[{\citenamefont{Maurya et~al.}(2016)\citenamefont{Maurya, Gupta, TT,
  and Rahaman}}]{maurya2016new}
\bibinfo{author}{\bibfnamefont{S.}~\bibnamefont{Maurya}},
  \bibinfo{author}{\bibfnamefont{Y.}~\bibnamefont{Gupta}},
  \bibinfo{author}{\bibfnamefont{S.}~\bibnamefont{TT}}, \bibnamefont{and}
  \bibinfo{author}{\bibfnamefont{F.}~\bibnamefont{Rahaman}},
  \bibinfo{journal}{The European Physical Journal A}
  \textbf{\bibinfo{volume}{52}}, \bibinfo{pages}{191} (\bibinfo{year}{2016}).

\bibitem[{\citenamefont{Lobo}(2006)}]{Lobo:2005uf}
\bibinfo{author}{\bibfnamefont{F.~S.~N.} \bibnamefont{Lobo}},
  \bibinfo{journal}{Class. Quant. Grav.} \textbf{\bibinfo{volume}{23}},
  \bibinfo{pages}{1525} (\bibinfo{year}{2006}), \eprint{gr-qc/0508115}.

\bibitem[{\citenamefont{Perlmutter
  et~al.}(1999)}]{SupernovaCosmologyProject:1998vns}
\bibinfo{author}{\bibfnamefont{S.}~\bibnamefont{Perlmutter}}
  \bibnamefont{et~al.} (\bibinfo{collaboration}{Supernova Cosmology Project}),
  \bibinfo{journal}{Astrophys. J.} \textbf{\bibinfo{volume}{517}},
  \bibinfo{pages}{565} (\bibinfo{year}{1999}), \eprint{astro-ph/9812133}.

\bibitem[{\citenamefont{Beltracchi and
  Gondolo}(2019)}]{beltracchi2019formation}
\bibinfo{author}{\bibfnamefont{P.}~\bibnamefont{Beltracchi}} \bibnamefont{and}
  \bibinfo{author}{\bibfnamefont{P.}~\bibnamefont{Gondolo}},
  \bibinfo{journal}{Physical Review D} \textbf{\bibinfo{volume}{99}},
  \bibinfo{pages}{044037} (\bibinfo{year}{2019}).

\bibitem[{\citenamefont{Chapline}(2004)}]{Chapline:2004jfp}
\bibinfo{author}{\bibfnamefont{G.}~\bibnamefont{Chapline}},
  \bibinfo{journal}{eConf} \textbf{\bibinfo{volume}{C041213}},
  \bibinfo{pages}{0205} (\bibinfo{year}{2004}), \eprint{astro-ph/0503200}.

\bibitem[{\citenamefont{Riess et~al.}(2019)\citenamefont{Riess, Casertano,
  Yuan, Macri, and Scolnic}}]{Riess:2019cxk}
\bibinfo{author}{\bibfnamefont{A.~G.} \bibnamefont{Riess}},
  \bibinfo{author}{\bibfnamefont{S.}~\bibnamefont{Casertano}},
  \bibinfo{author}{\bibfnamefont{W.}~\bibnamefont{Yuan}},
  \bibinfo{author}{\bibfnamefont{L.~M.} \bibnamefont{Macri}}, \bibnamefont{and}
  \bibinfo{author}{\bibfnamefont{D.}~\bibnamefont{Scolnic}},
  \bibinfo{journal}{Astrophys. J.} \textbf{\bibinfo{volume}{876}},
  \bibinfo{pages}{85} (\bibinfo{year}{2019}), \eprint{1903.07603}.

\bibitem[{\citenamefont{Sushkov}(2005)}]{sushkov2005wormholes}
\bibinfo{author}{\bibfnamefont{S.}~\bibnamefont{Sushkov}},
  \bibinfo{journal}{Physical Review D} \textbf{\bibinfo{volume}{71}},
  \bibinfo{pages}{043520} (\bibinfo{year}{2005}).

\bibitem[{\citenamefont{Bibi et~al.}(2016)\citenamefont{Bibi, Feroze, and
  Siddiqui}}]{bibi2016solution}
\bibinfo{author}{\bibfnamefont{R.}~\bibnamefont{Bibi}},
  \bibinfo{author}{\bibfnamefont{T.}~\bibnamefont{Feroze}}, \bibnamefont{and}
  \bibinfo{author}{\bibfnamefont{A.~A.} \bibnamefont{Siddiqui}},
  \bibinfo{journal}{Canadian Journal of Physics} \textbf{\bibinfo{volume}{94}},
  \bibinfo{pages}{758} (\bibinfo{year}{2016}).

\bibitem[{\citenamefont{Feng et~al.}(2008)\citenamefont{Feng, Wang, Abdalla,
  and Su}}]{feng2008observational}
\bibinfo{author}{\bibfnamefont{C.}~\bibnamefont{Feng}},
  \bibinfo{author}{\bibfnamefont{B.}~\bibnamefont{Wang}},
  \bibinfo{author}{\bibfnamefont{E.}~\bibnamefont{Abdalla}}, \bibnamefont{and}
  \bibinfo{author}{\bibfnamefont{R.}~\bibnamefont{Su}},
  \bibinfo{journal}{Physics Letters B} \textbf{\bibinfo{volume}{665}},
  \bibinfo{pages}{111} (\bibinfo{year}{2008}).

\bibitem[{\citenamefont{Gupta and Maurya}(2011)}]{gupta2011class}
\bibinfo{author}{\bibfnamefont{Y.}~\bibnamefont{Gupta}} \bibnamefont{and}
  \bibinfo{author}{\bibfnamefont{S.~K.} \bibnamefont{Maurya}},
  \bibinfo{journal}{Astrophysics and Space Science}
  \textbf{\bibinfo{volume}{331}}, \bibinfo{pages}{135} (\bibinfo{year}{2011}).

\bibitem[{\citenamefont{Kiess}(2012)}]{kiess2012exact}
\bibinfo{author}{\bibfnamefont{T.~E.} \bibnamefont{Kiess}},
  \bibinfo{journal}{Astrophysics and Space Science}
  \textbf{\bibinfo{volume}{339}}, \bibinfo{pages}{329} (\bibinfo{year}{2012}).

\bibitem[{\citenamefont{Takisa and Maharaj}(2013)}]{takisa2013some}
\bibinfo{author}{\bibfnamefont{P.~M.} \bibnamefont{Takisa}} \bibnamefont{and}
  \bibinfo{author}{\bibfnamefont{S.}~\bibnamefont{Maharaj}},
  \bibinfo{journal}{General Relativity and Gravitation}
  \textbf{\bibinfo{volume}{45}}, \bibinfo{pages}{1951} (\bibinfo{year}{2013}).

\bibitem[{\citenamefont{Malaver}(2017)}]{malaver2017new}
\bibinfo{author}{\bibfnamefont{M.}~\bibnamefont{Malaver}},
  \bibinfo{journal}{International Journal of Systems Science and Applied
  Mathematics} \textbf{\bibinfo{volume}{2}}, \bibinfo{pages}{93}
  (\bibinfo{year}{2017}).

\bibitem[{\citenamefont{Malaver}(2018)}]{malaver2018generalized}
\bibinfo{author}{\bibfnamefont{M.}~\bibnamefont{Malaver}},
  \bibinfo{journal}{World Scientific News} \textbf{\bibinfo{volume}{92}},
  \bibinfo{pages}{327} (\bibinfo{year}{2018}).

\bibitem[{\citenamefont{Sunzu et~al.}(2014)\citenamefont{Sunzu, Maharaj, and
  Ray}}]{sunzu2014quark}
\bibinfo{author}{\bibfnamefont{J.~M.} \bibnamefont{Sunzu}},
  \bibinfo{author}{\bibfnamefont{S.~D.} \bibnamefont{Maharaj}},
  \bibnamefont{and} \bibinfo{author}{\bibfnamefont{S.}~\bibnamefont{Ray}},
  \bibinfo{journal}{Astrophysics and Space Science}
  \textbf{\bibinfo{volume}{354}}, \bibinfo{pages}{517} (\bibinfo{year}{2014}).

\bibitem[{\citenamefont{Usov}(2004)}]{usov2004electric}
\bibinfo{author}{\bibfnamefont{V.~V.} \bibnamefont{Usov}},
  \bibinfo{journal}{Physical review D} \textbf{\bibinfo{volume}{70}},
  \bibinfo{pages}{067301} (\bibinfo{year}{2004}).

\bibitem[{\citenamefont{Glendenning}(2012)}]{glendenning2012compact}
\bibinfo{author}{\bibfnamefont{N.~K.} \bibnamefont{Glendenning}},
  \emph{\bibinfo{title}{Compact stars: Nuclear physics, particle physics and
  general relativity}} (\bibinfo{publisher}{Springer Science \& Business
  Media}, \bibinfo{year}{2012}).

\bibitem[{\citenamefont{Das and Ali}(2015)}]{das2015anisotropic}
\bibinfo{author}{\bibfnamefont{K.}~\bibnamefont{Das}} \bibnamefont{and}
  \bibinfo{author}{\bibfnamefont{N.}~\bibnamefont{Ali}},
  \bibinfo{journal}{Astrophysics and Space Science}
  \textbf{\bibinfo{volume}{356}}, \bibinfo{pages}{57} (\bibinfo{year}{2015}).

\bibitem[{\citenamefont{Ray et~al.}(2003)\citenamefont{Ray, Espindola,
  Malheiro, Lemos, and Zanchin}}]{ray2003electrically}
\bibinfo{author}{\bibfnamefont{S.}~\bibnamefont{Ray}},
  \bibinfo{author}{\bibfnamefont{A.~L.} \bibnamefont{Espindola}},
  \bibinfo{author}{\bibfnamefont{M.}~\bibnamefont{Malheiro}},
  \bibinfo{author}{\bibfnamefont{J.~P.} \bibnamefont{Lemos}}, \bibnamefont{and}
  \bibinfo{author}{\bibfnamefont{V.~T.} \bibnamefont{Zanchin}},
  \bibinfo{journal}{Physical Review D} \textbf{\bibinfo{volume}{68}},
  \bibinfo{pages}{084004} (\bibinfo{year}{2003}).

\bibitem[{\citenamefont{Pant}(2011)}]{pant2011some}
\bibinfo{author}{\bibfnamefont{N.}~\bibnamefont{Pant}},
  \bibinfo{journal}{Astrophysics and Space Science}
  \textbf{\bibinfo{volume}{331}}, \bibinfo{pages}{633} (\bibinfo{year}{2011}).

\bibitem[{\citenamefont{Pant et~al.}(2011)\citenamefont{Pant, Mehta, and
  Pant}}]{pant2011well}
\bibinfo{author}{\bibfnamefont{N.}~\bibnamefont{Pant}},
  \bibinfo{author}{\bibfnamefont{R.~N.} \bibnamefont{Mehta}}, \bibnamefont{and}
  \bibinfo{author}{\bibfnamefont{M.}~\bibnamefont{Pant}},
  \bibinfo{journal}{Astrophysics and Space Science}
  \textbf{\bibinfo{volume}{332}}, \bibinfo{pages}{473} (\bibinfo{year}{2011}).

\bibitem[{\citenamefont{Pant et~al.}(2019)\citenamefont{Pant, Gedela, Bisht,
  and Pant}}]{pant2019core}
\bibinfo{author}{\bibfnamefont{R.}~\bibnamefont{Pant}},
  \bibinfo{author}{\bibfnamefont{S.}~\bibnamefont{Gedela}},
  \bibinfo{author}{\bibfnamefont{R.~K.} \bibnamefont{Bisht}}, \bibnamefont{and}
  \bibinfo{author}{\bibfnamefont{N.}~\bibnamefont{Pant}}, \bibinfo{journal}{The
  European Physical Journal C} \textbf{\bibinfo{volume}{79}},
  \bibinfo{pages}{1} (\bibinfo{year}{2019}).

\bibitem[{\citenamefont{Pant et~al.}(2020)\citenamefont{Pant, Gedela, Pant,
  Upreti, and Bisht}}]{pant2020three}
\bibinfo{author}{\bibfnamefont{N.}~\bibnamefont{Pant}},
  \bibinfo{author}{\bibfnamefont{S.}~\bibnamefont{Gedela}},
  \bibinfo{author}{\bibfnamefont{R.}~\bibnamefont{Pant}},
  \bibinfo{author}{\bibfnamefont{J.}~\bibnamefont{Upreti}}, \bibnamefont{and}
  \bibinfo{author}{\bibfnamefont{R.~K.} \bibnamefont{Bisht}},
  \bibinfo{journal}{The European Physical Journal Plus}
  \textbf{\bibinfo{volume}{135}}, \bibinfo{pages}{1} (\bibinfo{year}{2020}).

\bibitem[{\citenamefont{Singh et~al.}(2017)\citenamefont{Singh, Pant, and
  Govender}}]{singh2017physical}
\bibinfo{author}{\bibfnamefont{K.~N.} \bibnamefont{Singh}},
  \bibinfo{author}{\bibfnamefont{N.}~\bibnamefont{Pant}}, \bibnamefont{and}
  \bibinfo{author}{\bibfnamefont{M.}~\bibnamefont{Govender}},
  \bibinfo{journal}{The European Physical Journal C}
  \textbf{\bibinfo{volume}{77}}, \bibinfo{pages}{1} (\bibinfo{year}{2017}).

\bibitem[{\citenamefont{Gedela et~al.}(2019)\citenamefont{Gedela, Pant, Upreti,
  and Pant}}]{gedela2019relativistic}
\bibinfo{author}{\bibfnamefont{S.}~\bibnamefont{Gedela}},
  \bibinfo{author}{\bibfnamefont{N.}~\bibnamefont{Pant}},
  \bibinfo{author}{\bibfnamefont{J.}~\bibnamefont{Upreti}}, \bibnamefont{and}
  \bibinfo{author}{\bibfnamefont{R.}~\bibnamefont{Pant}}, \bibinfo{journal}{The
  European Physical Journal C} \textbf{\bibinfo{volume}{79}},
  \bibinfo{pages}{1} (\bibinfo{year}{2019}).

\bibitem[{\citenamefont{Mafa~Takisa and Maharaj}(2013)}]{mafa2013compact}
\bibinfo{author}{\bibfnamefont{P.}~\bibnamefont{Mafa~Takisa}} \bibnamefont{and}
  \bibinfo{author}{\bibfnamefont{S.}~\bibnamefont{Maharaj}},
  \bibinfo{journal}{Astrophysics and Space Science}
  \textbf{\bibinfo{volume}{343}}, \bibinfo{pages}{569} (\bibinfo{year}{2013}).

\bibitem[{\citenamefont{Ngubelanga and Maharaj}(2015)}]{ngubelanga2015ray}
\bibinfo{author}{\bibfnamefont{S.}~\bibnamefont{Ngubelanga}} \bibnamefont{and}
  \bibinfo{author}{\bibfnamefont{S.}~\bibnamefont{Maharaj}},
  \bibinfo{journal}{Astrophys. Space Sci} p.~\bibinfo{pages}{74}
  (\bibinfo{year}{2015}).

\bibitem[{\citenamefont{Matondo et~al.}(2018)\citenamefont{Matondo, Maharaj,
  and Ray}}]{matondo2018relativistic}
\bibinfo{author}{\bibfnamefont{D.~K.} \bibnamefont{Matondo}},
  \bibinfo{author}{\bibfnamefont{S.}~\bibnamefont{Maharaj}}, \bibnamefont{and}
  \bibinfo{author}{\bibfnamefont{S.}~\bibnamefont{Ray}}, \bibinfo{journal}{The
  European Physical Journal C} \textbf{\bibinfo{volume}{78}},
  \bibinfo{pages}{1} (\bibinfo{year}{2018}).

\bibitem[{\citenamefont{Singh et~al.}(2020)\citenamefont{Singh, Ali, Rahaman,
  and Nasri}}]{singh2020compact}
\bibinfo{author}{\bibfnamefont{K.~N.} \bibnamefont{Singh}},
  \bibinfo{author}{\bibfnamefont{A.}~\bibnamefont{Ali}},
  \bibinfo{author}{\bibfnamefont{F.}~\bibnamefont{Rahaman}}, \bibnamefont{and}
  \bibinfo{author}{\bibfnamefont{S.}~\bibnamefont{Nasri}},
  \bibinfo{journal}{Physics of the Dark Universe}
  \textbf{\bibinfo{volume}{29}}, \bibinfo{pages}{100575}
  (\bibinfo{year}{2020}).

\bibitem[{\citenamefont{Rahaman et~al.}(2012)\citenamefont{Rahaman, Maulick,
  Yadav, Ray, and Sharma}}]{rahaman2012singularity}
\bibinfo{author}{\bibfnamefont{F.}~\bibnamefont{Rahaman}},
  \bibinfo{author}{\bibfnamefont{R.}~\bibnamefont{Maulick}},
  \bibinfo{author}{\bibfnamefont{A.~K.} \bibnamefont{Yadav}},
  \bibinfo{author}{\bibfnamefont{S.}~\bibnamefont{Ray}}, \bibnamefont{and}
  \bibinfo{author}{\bibfnamefont{R.}~\bibnamefont{Sharma}},
  \bibinfo{journal}{General Relativity and Gravitation}
  \textbf{\bibinfo{volume}{44}}, \bibinfo{pages}{107} (\bibinfo{year}{2012}).

\bibitem[{\citenamefont{Maharaj et~al.}(2014)\citenamefont{Maharaj, Sunzu, and
  Ray}}]{maharaj2014some}
\bibinfo{author}{\bibfnamefont{S.}~\bibnamefont{Maharaj}},
  \bibinfo{author}{\bibfnamefont{J.}~\bibnamefont{Sunzu}}, \bibnamefont{and}
  \bibinfo{author}{\bibfnamefont{S.}~\bibnamefont{Ray}}, \bibinfo{journal}{The
  European Physical Journal Plus} \textbf{\bibinfo{volume}{129}},
  \bibinfo{pages}{1} (\bibinfo{year}{2014}).

\bibitem[{\citenamefont{Estevez-Delgado and
  Estevez-Delgado}(2020)}]{estevez2020quintessence}
\bibinfo{author}{\bibfnamefont{G.}~\bibnamefont{Estevez-Delgado}}
  \bibnamefont{and}
  \bibinfo{author}{\bibfnamefont{J.}~\bibnamefont{Estevez-Delgado}},
  \bibinfo{journal}{The European Physical Journal C}
  \textbf{\bibinfo{volume}{80}}, \bibinfo{pages}{1} (\bibinfo{year}{2020}).

\bibitem[{\citenamefont{Estevez-Delgado
  et~al.}(2020)\citenamefont{Estevez-Delgado, Estevez-Delgado, Soto-Espitia,
  Duran, and Murgu{\'\i}a}}]{estevez2020tolman}
\bibinfo{author}{\bibfnamefont{G.}~\bibnamefont{Estevez-Delgado}},
  \bibinfo{author}{\bibfnamefont{J.}~\bibnamefont{Estevez-Delgado}},
  \bibinfo{author}{\bibfnamefont{R.}~\bibnamefont{Soto-Espitia}},
  \bibinfo{author}{\bibfnamefont{M.~P.} \bibnamefont{Duran}}, \bibnamefont{and}
  \bibinfo{author}{\bibfnamefont{A.~T.} \bibnamefont{Murgu{\'\i}a}},
  \bibinfo{journal}{The European Physical Journal Plus}
  \textbf{\bibinfo{volume}{135}}, \bibinfo{pages}{143} (\bibinfo{year}{2020}).

\bibitem[{\citenamefont{Chan et~al.}(2009{\natexlab{a}})\citenamefont{Chan,
  da~Silva, and Villas~da Rocha}}]{chan2009star}
\bibinfo{author}{\bibfnamefont{R.}~\bibnamefont{Chan}},
  \bibinfo{author}{\bibfnamefont{M.}~\bibnamefont{da~Silva}}, \bibnamefont{and}
  \bibinfo{author}{\bibfnamefont{J.~F.} \bibnamefont{Villas~da Rocha}},
  \bibinfo{journal}{General Relativity and Gravitation}
  \textbf{\bibinfo{volume}{41}}, \bibinfo{pages}{1835}
  (\bibinfo{year}{2009}{\natexlab{a}}).

\bibitem[{\citenamefont{Chan et~al.}(2009{\natexlab{b}})\citenamefont{Chan,
  Da~Silva, and Villas Da~Rocha}}]{chan2009anisotropic}
\bibinfo{author}{\bibfnamefont{R.}~\bibnamefont{Chan}},
  \bibinfo{author}{\bibfnamefont{M.}~\bibnamefont{Da~Silva}}, \bibnamefont{and}
  \bibinfo{author}{\bibfnamefont{J.~F.} \bibnamefont{Villas Da~Rocha}},
  \bibinfo{journal}{Modern Physics Letters A} \textbf{\bibinfo{volume}{24}},
  \bibinfo{pages}{1137} (\bibinfo{year}{2009}{\natexlab{b}}).

\bibitem[{\citenamefont{Lobo and Arellano}(2007)}]{lobo2007gravastars}
\bibinfo{author}{\bibfnamefont{F.~S.} \bibnamefont{Lobo}} \bibnamefont{and}
  \bibinfo{author}{\bibfnamefont{A.~V.} \bibnamefont{Arellano}},
  \bibinfo{journal}{Classical and Quantum Gravity}
  \textbf{\bibinfo{volume}{24}}, \bibinfo{pages}{1069} (\bibinfo{year}{2007}).

\bibitem[{\citenamefont{Chan et~al.}(2011)\citenamefont{Chan, Da~Silva, and
  Rocha}}]{chan2011gravastars}
\bibinfo{author}{\bibfnamefont{R.}~\bibnamefont{Chan}},
  \bibinfo{author}{\bibfnamefont{M.}~\bibnamefont{Da~Silva}}, \bibnamefont{and}
  \bibinfo{author}{\bibfnamefont{P.}~\bibnamefont{Rocha}},
  \bibinfo{journal}{General Relativity and Gravitation}
  \textbf{\bibinfo{volume}{43}}, \bibinfo{pages}{2223} (\bibinfo{year}{2011}).

\bibitem[{\citenamefont{Bertolami and Paramos}(2005)}]{bertolami2005chaplygin}
\bibinfo{author}{\bibfnamefont{O.}~\bibnamefont{Bertolami}} \bibnamefont{and}
  \bibinfo{author}{\bibfnamefont{J.}~\bibnamefont{Paramos}},
  \bibinfo{journal}{Physical Review D} \textbf{\bibinfo{volume}{72}},
  \bibinfo{pages}{123512} (\bibinfo{year}{2005}).

\bibitem[{\citenamefont{Cattoen and Faber}(2005)}]{cattoen2005visser}
\bibinfo{author}{\bibfnamefont{C.}~\bibnamefont{Cattoen}} \bibnamefont{and}
  \bibinfo{author}{\bibfnamefont{T.}~\bibnamefont{Faber}},
  \bibinfo{journal}{Class. Quantum Gravity} p. \bibinfo{pages}{4189}
  (\bibinfo{year}{2005}).

\bibitem[{\citenamefont{Bhar et~al.}(2018)\citenamefont{Bhar, Manna, Rahaman,
  and Banerjee}}]{bhar2018dark}
\bibinfo{author}{\bibfnamefont{P.}~\bibnamefont{Bhar}},
  \bibinfo{author}{\bibfnamefont{T.}~\bibnamefont{Manna}},
  \bibinfo{author}{\bibfnamefont{F.}~\bibnamefont{Rahaman}}, \bibnamefont{and}
  \bibinfo{author}{\bibfnamefont{A.}~\bibnamefont{Banerjee}},
  \bibinfo{journal}{Canadian Journal of Physics} \textbf{\bibinfo{volume}{96}},
  \bibinfo{pages}{594} (\bibinfo{year}{2018}).

\bibitem[{\citenamefont{Bhar}(2021)}]{bhar2021dark}
\bibinfo{author}{\bibfnamefont{P.}~\bibnamefont{Bhar}},
  \bibinfo{journal}{Physics of the Dark Universe}
  \textbf{\bibinfo{volume}{34}}, \bibinfo{pages}{100879}
  (\bibinfo{year}{2021}).

\bibitem[{\citenamefont{Ghezzi}(2011)}]{Ghezzi:2009ct}
\bibinfo{author}{\bibfnamefont{C.~R.} \bibnamefont{Ghezzi}},
  \bibinfo{journal}{Astrophys. Space Sci.} \textbf{\bibinfo{volume}{333}},
  \bibinfo{pages}{437} (\bibinfo{year}{2011}), \eprint{0908.0779}.

\bibitem[{\citenamefont{Tolman}(1939)}]{Tolman:1939jz}
\bibinfo{author}{\bibfnamefont{R.~C.} \bibnamefont{Tolman}},
  \bibinfo{journal}{Phys. Rev.} \textbf{\bibinfo{volume}{55}},
  \bibinfo{pages}{364} (\bibinfo{year}{1939}).

\bibitem[{\citenamefont{Kuchowicz}(1968)}]{osti_4507306}
\bibinfo{author}{\bibfnamefont{B.}~\bibnamefont{Kuchowicz}},
  \bibinfo{journal}{Acta Phys. Pol., 33: 541-63}  (\bibinfo{year}{1968}).

\bibitem[{\citenamefont{Ghezzi}(2005)}]{Ghezzi:2005iy}
\bibinfo{author}{\bibfnamefont{C.~R.} \bibnamefont{Ghezzi}},
  \bibinfo{journal}{Phys. Rev. D} \textbf{\bibinfo{volume}{72}},
  \bibinfo{pages}{104017} (\bibinfo{year}{2005}), \eprint{gr-qc/0510106}.

\bibitem[{\citenamefont{Barreto et~al.}(2007)\citenamefont{Barreto, Rodriguez,
  Rosales, and Serrano}}]{Barreto:2006cr}
\bibinfo{author}{\bibfnamefont{W.}~\bibnamefont{Barreto}},
  \bibinfo{author}{\bibfnamefont{B.}~\bibnamefont{Rodriguez}},
  \bibinfo{author}{\bibfnamefont{L.}~\bibnamefont{Rosales}}, \bibnamefont{and}
  \bibinfo{author}{\bibfnamefont{O.}~\bibnamefont{Serrano}},
  \bibinfo{journal}{Gen. Rel. Grav.} \textbf{\bibinfo{volume}{39}},
  \bibinfo{pages}{23} (\bibinfo{year}{2007}), \bibinfo{note}{[Erratum:
  Gen.Rel.Grav. 39, 537--538 (2007)]}, \eprint{gr-qc/0611089}.

\bibitem[{\citenamefont{Reissner}(1916)}]{reissner1916eigengravitation}
\bibinfo{author}{\bibfnamefont{H.}~\bibnamefont{Reissner}},
  \bibinfo{journal}{Annalen der Physik} \textbf{\bibinfo{volume}{355}},
  \bibinfo{pages}{106} (\bibinfo{year}{1916}).

\bibitem[{\citenamefont{Nordstr{\"o}m}(1918)}]{nordstrom1918energy}
\bibinfo{author}{\bibfnamefont{G.}~\bibnamefont{Nordstr{\"o}m}},
  \bibinfo{journal}{Koninklijke Nederlandse Akademie van Wetenschappen
  Proceedings Series B Physical Sciences} \textbf{\bibinfo{volume}{20}},
  \bibinfo{pages}{1238} (\bibinfo{year}{1918}).

\bibitem[{\citenamefont{Abubekerov et~al.}(2008)\citenamefont{Abubekerov,
  Antokhina, Cherepashchuk, and Shimanskii}}]{Abubekerov:2008inw}
\bibinfo{author}{\bibfnamefont{M.~K.} \bibnamefont{Abubekerov}},
  \bibinfo{author}{\bibfnamefont{E.~A.} \bibnamefont{Antokhina}},
  \bibinfo{author}{\bibfnamefont{A.~M.} \bibnamefont{Cherepashchuk}},
  \bibnamefont{and} \bibinfo{author}{\bibfnamefont{V.~V.}
  \bibnamefont{Shimanskii}}, \bibinfo{journal}{Astron. Rep.}
  \textbf{\bibinfo{volume}{52}}, \bibinfo{pages}{379} (\bibinfo{year}{2008}),
  \eprint{1201.5519}.

\bibitem[{\citenamefont{Varela et~al.}(2010)\citenamefont{Varela, Rahaman, Ray,
  Chakraborty, and Kalam}}]{Varela:2010mf}
\bibinfo{author}{\bibfnamefont{V.}~\bibnamefont{Varela}},
  \bibinfo{author}{\bibfnamefont{F.}~\bibnamefont{Rahaman}},
  \bibinfo{author}{\bibfnamefont{S.}~\bibnamefont{Ray}},
  \bibinfo{author}{\bibfnamefont{K.}~\bibnamefont{Chakraborty}},
  \bibnamefont{and} \bibinfo{author}{\bibfnamefont{M.}~\bibnamefont{Kalam}},
  \bibinfo{journal}{Phys. Rev. D} \textbf{\bibinfo{volume}{82}},
  \bibinfo{pages}{044052} (\bibinfo{year}{2010}), \eprint{1004.2165}.

\bibitem[{\citenamefont{Delgaty and Lake}(1998)}]{Delgaty:1998uy}
\bibinfo{author}{\bibfnamefont{M.~S.~R.} \bibnamefont{Delgaty}}
  \bibnamefont{and} \bibinfo{author}{\bibfnamefont{K.}~\bibnamefont{Lake}},
  \bibinfo{journal}{Comput. Phys. Commun.} \textbf{\bibinfo{volume}{115}},
  \bibinfo{pages}{395} (\bibinfo{year}{1998}), \eprint{gr-qc/9809013}.

\bibitem[{\citenamefont{Pant}(2010)}]{Pant:2010iub}
\bibinfo{author}{\bibfnamefont{N.}~\bibnamefont{Pant}},
  \bibinfo{journal}{Astrophys. Space Sci.} \textbf{\bibinfo{volume}{331}},
  \bibinfo{pages}{633} (\bibinfo{year}{2010}).

\bibitem[{\citenamefont{Chu and Tan}(2022)}]{Chu:2021uec}
\bibinfo{author}{\bibfnamefont{C.-S.} \bibnamefont{Chu}} \bibnamefont{and}
  \bibinfo{author}{\bibfnamefont{H.~S.} \bibnamefont{Tan}},
  \bibinfo{journal}{Universe} \textbf{\bibinfo{volume}{8}},
  \bibinfo{pages}{250} (\bibinfo{year}{2022}), \eprint{2103.06314}.

\bibitem[{\citenamefont{Darmois}(1927)}]{darmois1927equations}
\bibinfo{author}{\bibfnamefont{G.}~\bibnamefont{Darmois}},
  \bibinfo{journal}{Paris France}  (\bibinfo{year}{1927}).

\bibitem[{\citenamefont{Israel}(1966)}]{Israel:1966rt}
\bibinfo{author}{\bibfnamefont{W.}~\bibnamefont{Israel}},
  \bibinfo{journal}{Nuovo Cim. B} \textbf{\bibinfo{volume}{44S10}},
  \bibinfo{pages}{1} (\bibinfo{year}{1966}), \bibinfo{note}{[Erratum: Nuovo
  Cim.B 48, 463 (1967)]}.

\bibitem[{\citenamefont{Chandrasekhar}(1984)}]{chandrasekhar1984stars}
\bibinfo{author}{\bibfnamefont{S.}~\bibnamefont{Chandrasekhar}},
  \bibinfo{journal}{Science} \textbf{\bibinfo{volume}{226}},
  \bibinfo{pages}{497} (\bibinfo{year}{1984}).

\bibitem[{\citenamefont{Peebles and Ratra}(2003)}]{Peebles:2002gy}
\bibinfo{author}{\bibfnamefont{P.~J.~E.} \bibnamefont{Peebles}}
  \bibnamefont{and} \bibinfo{author}{\bibfnamefont{B.}~\bibnamefont{Ratra}},
  \bibinfo{journal}{Rev. Mod. Phys.} \textbf{\bibinfo{volume}{75}},
  \bibinfo{pages}{559} (\bibinfo{year}{2003}), \eprint{astro-ph/0207347}.

\bibitem[{\citenamefont{Baum and Frampton}(2007)}]{Baum:2006ee}
\bibinfo{author}{\bibfnamefont{L.}~\bibnamefont{Baum}} \bibnamefont{and}
  \bibinfo{author}{\bibfnamefont{P.~H.} \bibnamefont{Frampton}},
  \bibinfo{journal}{Phys. Rev. Lett.} \textbf{\bibinfo{volume}{98}},
  \bibinfo{pages}{071301} (\bibinfo{year}{2007}), \eprint{hep-th/0610213}.

\bibitem[{\citenamefont{Buchdahl}(1959)}]{buchdahl1959general}
\bibinfo{author}{\bibfnamefont{H.~A.} \bibnamefont{Buchdahl}},
  \bibinfo{journal}{Physical Review} \textbf{\bibinfo{volume}{116}},
  \bibinfo{pages}{1027} (\bibinfo{year}{1959}).

\bibitem[{\citenamefont{Florides}(1983)}]{florides1983complete}
\bibinfo{author}{\bibfnamefont{P.~S.} \bibnamefont{Florides}},
  \bibinfo{journal}{Journal of Physics A: Mathematical and General}
  \textbf{\bibinfo{volume}{16}}, \bibinfo{pages}{1419} (\bibinfo{year}{1983}).

\bibitem[{\citenamefont{Kumar and Bharti}(2022)}]{kumar2022isotropic}
\bibinfo{author}{\bibfnamefont{J.}~\bibnamefont{Kumar}} \bibnamefont{and}
  \bibinfo{author}{\bibfnamefont{P.}~\bibnamefont{Bharti}},
  \bibinfo{journal}{The European Physical Journal Plus}
  \textbf{\bibinfo{volume}{137}}, \bibinfo{pages}{330} (\bibinfo{year}{2022}).

\bibitem[{\citenamefont{Boehmer and Harko}(2007)}]{Boehmer:2007gq}
\bibinfo{author}{\bibfnamefont{C.~G.} \bibnamefont{Boehmer}} \bibnamefont{and}
  \bibinfo{author}{\bibfnamefont{T.}~\bibnamefont{Harko}},
  \bibinfo{journal}{Gen. Rel. Grav.} \textbf{\bibinfo{volume}{39}},
  \bibinfo{pages}{757} (\bibinfo{year}{2007}), \eprint{gr-qc/0702078}.

\bibitem[{\citenamefont{Andreasson}(2009)}]{Andreasson:2008xw}
\bibinfo{author}{\bibfnamefont{H.}~\bibnamefont{Andreasson}},
  \bibinfo{journal}{Commun. Math. Phys.} \textbf{\bibinfo{volume}{288}},
  \bibinfo{pages}{715} (\bibinfo{year}{2009}), \eprint{0804.1882}.

\bibitem[{\citenamefont{Herrera}(1992)}]{Herrera:1992lwz}
\bibinfo{author}{\bibfnamefont{L.}~\bibnamefont{Herrera}},
  \bibinfo{journal}{Phys. Lett. A} \textbf{\bibinfo{volume}{165}},
  \bibinfo{pages}{206} (\bibinfo{year}{1992}).

\bibitem[{\citenamefont{Abreu et~al.}(2007)\citenamefont{Abreu, Hernandez, and
  Nunez}}]{Abreu:2007ew}
\bibinfo{author}{\bibfnamefont{H.}~\bibnamefont{Abreu}},
  \bibinfo{author}{\bibfnamefont{H.}~\bibnamefont{Hernandez}},
  \bibnamefont{and} \bibinfo{author}{\bibfnamefont{L.~A.} \bibnamefont{Nunez}},
  \bibinfo{journal}{Class. Quant. Grav.} \textbf{\bibinfo{volume}{24}},
  \bibinfo{pages}{4631} (\bibinfo{year}{2007}), \eprint{0706.3452}.

\bibitem[{\citenamefont{Bondi}(1947)}]{bondi1947spherically}
\bibinfo{author}{\bibfnamefont{H.}~\bibnamefont{Bondi}},
  \bibinfo{journal}{Monthly Notices of the Royal Astronomical Society}
  \textbf{\bibinfo{volume}{107}}, \bibinfo{pages}{410} (\bibinfo{year}{1947}).

\bibitem[{\citenamefont{Witten}(1981)}]{witten1981new}
\bibinfo{author}{\bibfnamefont{E.}~\bibnamefont{Witten}},
  \bibinfo{journal}{Communications in Mathematical Physics}
  \textbf{\bibinfo{volume}{80}}, \bibinfo{pages}{381} (\bibinfo{year}{1981}).

\bibitem[{\citenamefont{Visser}(1997)}]{visser1997energy}
\bibinfo{author}{\bibfnamefont{M.}~\bibnamefont{Visser}},
  \bibinfo{journal}{Science} \textbf{\bibinfo{volume}{276}},
  \bibinfo{pages}{88} (\bibinfo{year}{1997}).

\bibitem[{\citenamefont{Garcia et~al.}(2011)\citenamefont{Garcia, Harko, Lobo,
  and Mimoso}}]{garcia2011energy}
\bibinfo{author}{\bibfnamefont{N.~M.} \bibnamefont{Garcia}},
  \bibinfo{author}{\bibfnamefont{T.}~\bibnamefont{Harko}},
  \bibinfo{author}{\bibfnamefont{F.~S.} \bibnamefont{Lobo}}, \bibnamefont{and}
  \bibinfo{author}{\bibfnamefont{J.~P.} \bibnamefont{Mimoso}},
  \bibinfo{journal}{Physical Review D} \textbf{\bibinfo{volume}{83}},
  \bibinfo{pages}{104032} (\bibinfo{year}{2011}).

\bibitem[{\citenamefont{Ponce~de Leon}(1993)}]{ponce1993limiting}
\bibinfo{author}{\bibfnamefont{J.}~\bibnamefont{Ponce~de Leon}},
  \bibinfo{journal}{General relativity and gravitation}
  \textbf{\bibinfo{volume}{25}}, \bibinfo{pages}{1123} (\bibinfo{year}{1993}).

\bibitem[{\citenamefont{Tolman}(1930)}]{PhysRev.35.875}
\bibinfo{author}{\bibfnamefont{R.~C.} \bibnamefont{Tolman}},
  \bibinfo{journal}{Phys. Rev.} \textbf{\bibinfo{volume}{35}},
  \bibinfo{pages}{875} (\bibinfo{year}{1930}),
  \urlprefix\url{https://link.aps.org/doi/10.1103/PhysRev.35.875}.

\bibitem[{\citenamefont{Shapiro and Teukolsky}(2008)}]{shapiro2008black}
\bibinfo{author}{\bibfnamefont{S.~L.} \bibnamefont{Shapiro}} \bibnamefont{and}
  \bibinfo{author}{\bibfnamefont{S.~A.} \bibnamefont{Teukolsky}},
  \emph{\bibinfo{title}{Black holes, white dwarfs, and neutron stars: The
  physics of compact objects}} (\bibinfo{publisher}{John Wiley \& Sons},
  \bibinfo{year}{2008}).

\bibitem[{\citenamefont{Zeldovich and
  Novikov}(1971)}]{zeldovich1971relativistic}
\bibinfo{author}{\bibfnamefont{Y.~B.} \bibnamefont{Zeldovich}}
  \bibnamefont{and} \bibinfo{author}{\bibfnamefont{I.~D.}
  \bibnamefont{Novikov}}, \bibinfo{journal}{Chicago: University of Chicago
  Press}  (\bibinfo{year}{1971}).

\bibitem[{\citenamefont{Zel'dovich et~al.}(1972)\citenamefont{Zel'dovich,
  Novikov, and Silk}}]{zel1972relativistic}
\bibinfo{author}{\bibfnamefont{Y.~B.} \bibnamefont{Zel'dovich}},
  \bibinfo{author}{\bibfnamefont{I.~D.} \bibnamefont{Novikov}},
  \bibnamefont{and} \bibinfo{author}{\bibfnamefont{J.}~\bibnamefont{Silk}},
  \bibinfo{journal}{Physics Today} \textbf{\bibinfo{volume}{25}},
  \bibinfo{pages}{63} (\bibinfo{year}{1972}).

\bibitem[{\citenamefont{L'DOVICH}(1962)}]{l1962equation}
\bibinfo{author}{\bibfnamefont{Y.}~\bibnamefont{L'DOVICH}},
  \bibinfo{journal}{Sov. Phys. JETP} \textbf{\bibinfo{volume}{14}},
  \bibinfo{pages}{1609} (\bibinfo{year}{1962}).

\bibitem[{\citenamefont{Harrison et~al.}(1965)\citenamefont{Harrison, Thorne,
  Wakano, and Wheeler}}]{harrison1965gravitation}
\bibinfo{author}{\bibfnamefont{B.~K.} \bibnamefont{Harrison}},
  \bibinfo{author}{\bibfnamefont{K.~S.} \bibnamefont{Thorne}},
  \bibinfo{author}{\bibfnamefont{M.}~\bibnamefont{Wakano}}, \bibnamefont{and}
  \bibinfo{author}{\bibfnamefont{J.~A.} \bibnamefont{Wheeler}},
  \bibinfo{journal}{Gravitation Theory and Gravitational Collapse}
  (\bibinfo{year}{1965}).

\bibitem[{\citenamefont{{Shvartsman}}(1971)}]{1971JETP...33..475S}
\bibinfo{author}{\bibfnamefont{V.~F.} \bibnamefont{{Shvartsman}}},
  \bibinfo{journal}{Soviet Journal of Experimental and Theoretical Physics}
  \textbf{\bibinfo{volume}{33}}, \bibinfo{pages}{475} (\bibinfo{year}{1971}).

\end{thebibliography}

\end{document}